\documentclass[12pt,fleqn,twoside]{article}
\usepackage{graphicx}
\usepackage{latexsym}
\usepackage{fancyhdr}
\usepackage{cite}

\begin{document}
\pagenumbering{Roman}
\thispagestyle{empty}

\pagenumbering{Roman}
\begin{center}
\begin{large}
\textbf{On the Theory of Generalized Algebraic Transformations} \\
\end{large}
\begin{large}
Algebraic Mapping Transformations and Exactly Solvable \\ Models in Statistical Mechanics \\
\end{large}
\vspace{0.4cm}
\begin{large}
\textbf{Jozef Stre\v{c}ka}  \\
(habilitation thesis) 
\end{large}
\end{center}


\tableofcontents
\newpage

\pagestyle{fancy}
\pagenumbering{arabic}
\section{Introduction}

Statistical physics is fundamental physical theory, which deals with equilibrium (or even non-equilibrium) properties of a large number of particles using the well established 
concept based either on classical or quantum mechanics. With respect to this, the term statistical mechanics is often used as a synonym to statistical physics that covers probabilistic (statistical) approach to classical or quantum mechanics concerning with many-particle systems. The most important benefit resulting from this theory consists in that it relates microscopic properties of individual particles to observable macroscopic (bulk) properties of matter. Even although relations between some macroscopic properties and fundamental properties of individual particles are occasionally elementary (for instance the total mass is simply a sum over particle masses), many material properties cannot be simply elucidated from the fundamental properties of constituent particles, i.e., from the microscopic point of view. In particular, statistical mechanics enables to explain observable macroscopic 
features of real materials solely by imposing forces between the constituent particles. 
For this purpose, one necessarily needs just some plausible assumption about internal forces between constituent particles (inter-particle interactions) in order to make relevant theoretical predictions 
for observable properties of a given macroscopic system. This assumption, which is built on some realistic microscopic idea of how individual particles interact among themselves, constitutes a framework for some simple theoretical idealization to be referred to as a statistical model.
  
Of course, each \textit{statistical model} serves only as an approximative description of physical reality aimed at describing observable macroscopic properties preferably quantitatively or leastwise qualitatively. However, it is very difficult and often incredible task to define a realistic model, which is on the one hand mathematically tractable and on the other hand provides a comprehensive description of all observable macroscopic properties. The most formidable difficulties are usually encountered when attempting to formulate and to solve the relevant model mathematically. There are 
just few valuable exceptions. The most common example surely represents an exactly solvable model 
of an ideal gas (no matter whether consisting of classical particles or fermions or bosons)
in which the constituent particles do not interact among themselves until they undergo perfectly elastic collisions. If the inter-particle interactions are taken into account (suppose for instance 
the real gas instead of the ideal gas), however, realistic models are highly appreciated 
if they are still exactly solvable, but this is usually not the case. 

If the constituent particles of some interacting many-particle system are situated on discrete sites 
of a crystal lattice and only short-ranged inter-particle interactions need to be considered, 
then a substantial simplification in the mathematical treatment of relevant model(s) is 
usually achieved. Under these simplifying constraints, one concerns with so-called \textit{lattice-statistical models} that are generally more amenable to an exact analytical treatment even though sophisticated mathematical methods must be still employed for obtaining exact solutions 
of even relatively simple-minded models. Hence, it follows that exactly solved models are usually considered as an inspiring research field to emerge in the statistical mechanics, which regrettably requires a considerable knowledge of sophisticated mathematics and are therefore beyond the scope of standard courses on the statistical physics. Apart from this drawback, the topic \textit{exactly solvable models in statistical mechanics} surely represent an exciting research field in its own 
right as convincingly evidenced by a rather rich and instantly growing list of excellent books 
devoted to this intriguing subject matter  \cite{domb72,thom79,baxt82,matt93,stan93,king96,lavi99,yeom02,tana02,lieb04,suth04,diep04,wu09}. 

The main goal of this book is to make a brief introduction into the method of algebraic mapping transformations, i.e. an exact mathematical technique, which enables after relatively modest calculation a rigorous analytical treatment of diverse more complex lattice-statistical models 
by establishing a precise mapping correspondence with simpler exactly solved lattice-statistical models.

\newpage 

\section{Ising and Heisenberg models}

In this section, let make few comments on two generic lattice-statistical models, 
which are of particular research interest because of their usefulness and flexibility 
in representing diverse real-world systems of very different nature. 

\subsection{Ising model}

The Ising model perhaps represent the most versatile model of statistical mechanics at all, which is simultaneously fully mathematically tractable on one- and two-dimensional (1D and 2D) lattices. 
This simple-minded lattice-statistical model has been proposed by Lenz in 1920 \cite{lenz20} 
and five years later has been exactly solved by Ising for the particular case of 
the linear chain \cite{isin25}. Note furthermore that there are several 
excellent review articles on the historical developments of the Ising model to which 
the interested reader is referred to for further details \cite{brus67,kobe00,niss05,niss09}.

The spin-1/2 Ising model can be defined on a crystal lattice\footnote{Note that the Ising model 
can be defined for systems without translational invariance as well.} through the Hamiltonian
\begin{eqnarray}
{\cal H} = - J \sum_{(i,j)} \sigma_i \sigma_j - H \sum_{i=1}^N \sigma_i ,
\label{m1}	
\end{eqnarray}
where $\sigma_i = \pm 1/2$ is two-valued Ising spin variable situated at the $i$th site of 
a crystal lattice, the former summation takes into account a configurational energy associated 
with the interaction between the nearest-neighbour spins and the latter summation accounts for 
the Zeeman's energy $H = g \mu_B B$ of magnetic moments in an external magnetic field $B$ ($g$ 
stands for Land\'e $g$-factor and $\mu_B$ is Bohr magneton). At first sight, the Ising model 
might seem to be the greatly oversimplified model as it first takes into consideration only extremely short-ranged interactions\footnote{The first summation is usually restricted just to the pairs of nearest-neighbour spins.} and second, it also neglects all quantum effects in that it disregards 
a quantum-mechanical nature of spin by considering spin as a classical two-valued variable. 
While the former restriction is rather well satisfied in a variety of insulating magnetic 
materials, the latter restriction turns out to be much more profound as far as the theoretical 
modeling of insulating magnetic materials is concerned. From this point of view, the Ising model 
offers merely a semi-classical description of interacting many-particle systems, because a set 
of discrete spin values is the only quantum feature of this lattice-statistical model.      

Before proceeding to a survey of the exactly solved Ising models, let us briefly comment on 
their possible experimental realizations. The Ising model was for many years merely regarded 
as the purely academic model without any correspondence to a specific real-world system, 
since the first insulating magnetic materials
that would satisfy its specific requirements have been discovered almost a half century after 
its invention. At present, there are two wide families of insulating magnetic materials whose 
magnetic behaviour is generally in accord with theoretical predictions of the Ising model. 
The first class of the Ising-like magnetic materials involve \textit{rare-earth compounds} 
such as Dy(C$_2$H$_5$SO$_4$)$_3$.9H$_2$O, Dy$_3$Al$_5$O$_{12}$, DyPO$_4$, LiHoF$_4$, LiTbF$_4$ 
and so on (see Ref. \cite{wolf00} and references cited therein). 
In this class of materials, the only carriers of magnetic moments are the rare-earth elements 
(Dy, Ho, Tb, etc.) that interact among themselves almost exclusively through dipolar forces, 
while other non-dipolar interactions are in general negligible. Under certain conditions, the dipole-dipole interaction between the more distant magnetic moments can be ignored and it is sufficient 
to consider merely the interaction between the nearest-neighbour magnetic moments (Ising-like criterion). It should be nevertheless pointed out that the magnetic dipole-dipole interaction 
is long-range interaction in its character (it decays as an inverse third power of the distance) 
and hence, the interactions with more distant rare-earth ions can occasionally be at an origin 
of more complex behaviour. Thus, the most suitable experimental realizations of the Ising-like models 
provide rare-earth compounds in which the interactions between more distant magnetic moments almost completely cancel out and the interaction between the nearest-neighbour magnetic moments makes 
the most profound contribution to the overall magnetic behaviour. 

The second class of the Ising-like magnetic materials represent the insulating magnetic materials 
from the family of \textit{molecular-based compounds} \cite{jong74}. In this wide family of magnetic compounds, the most important interaction between magnetic centers (mostly transition-metal ions) 
is a pairwise spin-spin interaction mediated via intervening non-magnetic atom(s) through 
the indirect \textit{superexchange mechanism} \cite{kram34,ande50,ande59}. Even though there 
does not exist a general theory, which would admit a straightforward calculation of the interaction parameter $J$ originating from the superexchange mechanism and this parameter is usually obtained only as a self-adjustable parameter from a comparison with relevant experimental data, a strength of the spin-spin exchange interaction decays very rapidly with a distance - at least as an inverse 
tenth power of distance \cite{bloc66}. From this point of view, the molecular-based compounds 
involving magnetic metal centers satisfy much better the necessary criterion for the Ising-like materials, which demands a predominant nearest-neighbour interaction and negligible interactions between the more distant spins. The most crucial limitation, 
which prevents the most of molecular-based compounds to be good examples of the Ising-like materials, 
thus lies in a demand of having the extremely anisotropic exchange interaction while 
the superexchange mechanism gives rise to the isotropic exchange interaction \cite{jong74}. 
However, the anisotropic exchange interaction need not arise from the interaction mechanism alone, 
but it may have a close connection with another sources of the magnetic anisotropy such as spin-orbit coupling, crystal-field effect, dipolar interactions, etc \cite{wolf00,jong74}. Under these circumstances, the theoretical description based on the Ising model 
is justified even if it still represents a certain oversimplification of the physical reality. 
It is noteworthy that the overall agreement between theoretical predictions derived from the Ising model and the relevant experimental data is generally found to be very satisfactory mainly 
for several cobalt-based compounds such as Co(pyridine)$_2$Cl$_2$, A$_2$CoF$_4$ (A = K, Rb), Cs$_3$CoX$_5$ (X = F, Cl, Br) \cite{wolf00,jong74}.

Finally, it should be also remarked that all magnetic compounds from both the aforementioned families of the Ising-like materials are three-dimensional crystals in reality, however, some of them can effectively possess the \textit{low-dimensional magnetic structure} on behalf of the lack of an appreciable magnetic interaction in one or more spatial directions. As a matter of fact, the magnetic and crystallographic lattices may significantly differ one from each other especially if the carriers of magnetic moment are well separated along some spatial direction(s). Consequently, the magnetic lattice then becomes low-dimensional due to the short-range character of magnetic forces and it is therefore of fundamental importance to investigate low-dimensional spin models as well. The reliability of exactly solved low-dimensional Ising models in representing real-world insulating magnetic materials has been checked with an appreciable success even if some healthy skepticism is always appropriate if one is seeking true understanding of real magnetic materials \cite{wolf00}.

In conclusion, let us also briefly mention other possible (non-magnetic) applications of 
the Ising model and its different generalizations like the Blume-Capel model \cite{blum66,cape66}, 
the Blume-Emery-Griffiths model \cite{blum71} and others, which provide a deeper understanding of cooperative behaviour inherent to interacting many-particle systems from seemingly diverse research areas. Even though the Ising model has been originally invented for describing cooperative 
nature of spontaneous long-range order to emerge in the magnetic materials, throughout 
the years it has proved its usefulness by investigating the order-disorder phenomena 
in metal alloys \cite{yang52,lee52,bern82}, the vapour-liquid coexistence curves at 
liquid-gas transition \cite{yang52,lee52,whee77}, the phase separation in liquid mixtures \cite{blum71,whee77,whee68}, the saturation curve of hemoglobine \cite{thom79,mono65,thom68}, 
the initial reaction rate of allosteric enzymes \cite{thom68,thom79a}, the melting curve 
of helix-coil transition in DNA \cite{mont66,goel68,wart72,wart85}, the role of socio-economic interactions in determining business confidence indicators \cite{hohn05,stau06,stau08},
urban segregation \cite{sche71,stau07,mull08}, the language change \cite{stau08,itoh04,schu08} 
and many other topics.

\subsection{Survey of exactly solved Ising models}

A great deal of research interest aimed at searching various exactly solvable Ising models 
has resulted in a rather extensive list of up to date available literature concerned with 
the exactly solved Ising models. It is therefore beyond the scope of this work to review 
all of them and selected examples should mainly serve only for illustration and are chosen 
so as to reflect author's previous and current research interests. It should be also emphasized, moreover, that the exact solution of any non-planar Ising model is essentially NP-complete problem 
as recently pointed out by Istrail \cite{istr00} and Cipra \cite{cipr00}, which implies that 
any 3D Ising model is possibly analytically intractable problem despite its conceptual simplicity. 
Hence, the subsequent list of exactly solved Ising models essentially contains 
only the lattice-statistical models defined on 1D and 2D lattices.   

It has been already mentioned previously that the spin-1/2 Ising model on the linear chain 
has exactly been solved by Ising in 1925 with the help of combinatorial approach \cite{isin25}. 
Afterwards, the Ising's exact results have been re-derived using a variety of other mathematical 
techniques such as the transfer-matrix method due to Kramers and Wannier \cite{kram41}. 
Among the most interesting rigorously solved 1D Ising models one could mention the spin-$S$ 
Ising linear chain \cite{suzu67,line79,chat84}, the spin-1/2 Ising chain accounting for the interactions 
between the more distant spins \cite{step70,step71,dobs69}, the spin-1/2 Ising model defined 
for two, three or four coupled chains \cite{inou71,kalo75,fedr76,yoko89,yuri07}, the spin-1/2 
Ising model with alternating bonds \cite{ohan03} and the mixed-spin Ising chains \cite{urum80,geor85,cure86,cure92,fire97,fire03} and ladders \cite{geor92,kiss08} supplemented by 
the zero-field splitting parameters, biquadratic interaction, etc. It is worthwhile to remark 
that Minami \cite{mina96,mina98a,mina98b} has recently succeeded in obtaining a quite general 
exact solution for a rather large class of 1D Ising models, which involve the most 
of aforelisted exactly solved models including the ones with the mixed spins and alternating bonds. 

2D Ising model had resisted almost two decades of intensive efforts until the complete closed-form 
exact solution has been found by Onsager for the spin-1/2 Ising model on a square lattice without 
the external magnetic field using the transfer-matrix method and Lie algebra \cite{onsa44}. Onsager's 
exact solution is currently regarded as one of the most significant achievements in the equilibrium statistical mechanics, because it brought an important revision in the understanding of phase transitions and critical phenomena. Actually, Onsager's  exact results had served in evidence of a striking phase transition, which comes from extremely short-ranged interactions and is accompanied with a strange singular behaviour 
of several thermodynamic quantities in a close vicinity of the critical point. From this point 
of view, Onsager's exact solution also brought a considerable insight into deficiencies of some approximative methods, which mostly fail in predicting correct behaviour near a critical region. In this regard, one of the most essential questions to deal with in the statistical mechanics of exactly solvable models is always to find a precise nature of discontinuities and singularities accompanying each phase transition. 

The only disadvantage of Onsager's method lies in a considerable mathematical formida\-bility 
of his solution. Bearing this in mind, many theoretical physicists have started to search for 
an alternative way, which would admit a more straightforward exact treatment of 2D 
Ising models. It is noteworthy that the original Onsager's solution has been slightly simplified 
by Kaufmann and Onsager himself using the theory of spinors \cite{kauf49,onsa49}. However, 
the more substantial simplification has been later on achieved using various rigorous methods
such as several combinatorial approaches developed by Kac, Ward and Potts \cite{kac52,pott55}, 
Hurst and Green \cite{hurs60}, Vdovichenko \cite{vdov64,vdov65a,vdov65b} and others, 
the formalism of second quantization based on Jordan-Wigner fermionization invented 
by Schultz, Lieb and Mattis \cite{schu64}, 
the recurrence relations derived from the star-triangle transformation by Baxter and Enting \cite{baxt78}, or more recent theories based on Grassmann variables \cite{samu80,plec85,plec88,noji98} or Clifford-Dirac algebra \cite{verg09}. Among all rigorous 
techniques, the Pfaffian method seems to be the simplest method that enables to solve exactly 
the spin-1/2 Ising model on 2D lattices. Within this rigorous approach, 
the problem of solving 2D Ising model is firstly reformulated 
as the problem of dimer statistics on a relevant decorated lattice \cite{kast63} and 
subsequently, the relevant dimer statistics is precisely solved by employing the 
Pfaffian technique following the ideas of Kasteleyn \cite{kast61}, Temperley and Fisher \cite{temp61,fish61}. Note furthermore this exact method can be used for treating the spin-1/2 Ising model on any planar lattice without crossing bonds and among other matters, this method has thus proved a strong universality in a critical behaviour of 2D Ising lattices \cite{hurs63,gree64,hurs65,mcco73}. 

\subsection{Heisenberg model}

It has been already mentioned in the preceding part that the quantum Heisenberg model \cite{heis28} 
and its various extensions are much sought after, since they are more appropriate for modeling 
the magnetic behaviour of the most of real insulating magnetic materials than the aforedescribed Ising models. 
The main reason for a greater success of the Heisenberg-type models lies in the fact that 
these quantum spin models correctly take into account quantum fluctuations, which might play 
a crucial role in determining magnetic properties especially at low enough temperatures. 
In accordance with this statement, the magnetic behaviour of a rather extensive number of molecular-based magnetic compounds obeys theoretical predictions of the Heisenberg-type 
models as convincingly evidenced in the review article of de Jongh and Miedema \cite{jong74}, 
several earlier books devoted exclusively to the molecular-based magnetic materials \cite{will85,carl86,gatt91,kahn93,coro96,turn96}, as well as, the more recent series 
of books edited by Miller and Drillon \cite{mill01,mill02,mill03,mill04,mill05}. 
It is safety to say that it is even not possible to enumerate manifold application 
of the Heisenberg-type models in explaining magnetic properties of real molecular-based 
magnetic compounds and hence, we will not dwell further on this aspect with respect 
to a rather extensive list of excellent literature \cite{jong74,will85,carl86,gatt91,kahn93,coro96,turn96,mill01,mill02,mill03,mill04,mill05} 
to be published on this topic.

The quantum Heisenberg model can be defined through the Hamiltonian 
\begin{eqnarray}
\hat{{\cal H}} = - J \sum_{(i,j)} \vec{S}_i \cdot \vec{S}_j - H \sum_{i=1}^N \hat{S}_i^z ,
\label{h1}	
\end{eqnarray}
where the first summation usually runs over all pairs of nearest-neighbour spins, the second 
summation is carried out over all lattice sites and the symbol $\vec{S}_j$ marks the spin operator 
for the $j$th lattice site with the spatial components $\vec{S}_j \equiv (\hat S_j^x, \hat S_j^y, 
\hat S_j^z)$ to be given by two-by-two Pauli spin matrices\footnote{Here and in what follows the Pauli matrices are written in units $\hbar=1$.} 
\begin{eqnarray}
\hat{S}_j^x = \frac{1}{2} \left(
\begin{array}{cc}
0 & 1 \\
1 & 0	
\end{array} \right)_{\! \!j}, \quad 
\hat{S}_j^y = \frac{1}{2} \left(
\begin{array}{cc}
0 & -i \\
i & 0	\\
\end{array} \right)_{\! \! j}, \quad
\hat{S}_j^z = \frac{1}{2} \left(
\begin{array}{cc}
1 & 0 \\
0 & -1 \\	
\end{array} \right)_{\! \! j},
\label{h2}
\end{eqnarray} 
where $i = \sqrt{-1}$ is the imaginary unit. As before, the former term in the Hamiltonian (\ref{h1}) takes into consideration a configurational energy associated with the exchange interaction between the nearest-neighbour spins and the latter term accounts for the Zeeman's energy of discrete magnetic moments. In an attempt to treat exactly the quantum Heisenberg model one encounters the most crucial mathematical difficulties in a non-commutability of the spin operators, 
which obey the following set of commutation relations   
\begin{eqnarray}
[\hat{S}_j^{\alpha} , \hat{S}_k^{\beta} ] = \hat{S}_j^{\alpha} \hat{S}_k^{\beta} 
- \hat{S}_k^{\beta} \hat{S}_j^{\alpha} = i \hat{S}_j^{\gamma} \delta_{jk} 
\varepsilon_{\alpha \beta \gamma}, 
\label{h3}	
\end{eqnarray}
where $\alpha, \beta, \gamma \in \{ x,y,z \}$, the symbols $\delta_{jk}$ and $\varepsilon_{\alpha \beta \gamma}$ label Kronecker and Levi-Civita symbols, respectively. The most wide-spread extension of the quantum Heisenberg model (\ref{h1}) certainly represents the XXZ Heisenberg model given by the Hamiltonian
\begin{eqnarray}
\hat{{\cal H}} = - J \sum_{(i,j)} [ \Delta (\hat{S}_i^x \hat{S}_j^x + \hat{S}_i^y \hat{S}_j^y) 
+ \hat{S}_i^z \hat{S}_j^z] - H \sum_{i=1}^N \hat{S}_i^z,
\label{h4}	
\end{eqnarray}
which merely replaces the isotropic pairwise spin-spin interaction through the anisotropic one. 
The anisotropy parameter $\Delta$ then allows to obtain the semi-classical Ising model 
as a special limiting case of the XXZ Heisenberg model in the limit $\Delta \to 0$. 
Another special limiting case of the XXZ Heisenberg model, which is of particular research 
interest, represents the quantum XY model obtained from the Hamiltonian (\ref{h4}) 
by considering the other extreme limiting case $\Delta \to \infty$.

\subsection{Survey of exactly solved Heisenberg models}

Owing to mathematical complexities, which are closely connected with a non-commutabi\-lity of the 
spin operators, the list of exactly solved quantum Heisenberg models is much less numerous 
compared to the above mentioned list of exactly solved Ising models. It is worthy of notice 
that the exactly solved quantum Heisenberg models essentially comprise quantum spin models, which 
have been exactly solved only at zero temperature (ground state). Moreover, it should be also 
stressed that the quantum Heisenberg models exhibit a more diverse behaviour compared to 
the semi-classical Ising models just on behalf of the non-commutability of the spin operators, 
which in turn causes a presence of quantum fluctuations fundamentally influencing 
the ground state of the antiferromagnetic Heisenberg models. From this point of view, 
the ferromagnetic and antiferromagnetic quantum Heisenberg models differ basically one from another
in their behaviour, the former exhibits a classical ferromagnetic ground state, whereas the latter 
one often exhibits a variety of unusual and exotic ground state(s) as the classical antiferromagnetic 
(N\'eel) order is not a true eigenstate of the Hamiltonians (\ref{h1}) and (\ref{h4}). 

Bearing all this in mind, the antiferromagnetic quantum Heisenberg models remain at the forefront of theoretical research interest over the past few decades. The eigenstates and the ground-state energy 
of the antiferromagnetic spin-1/2 Heisenberg linear chain has been exactly found by Bethe \cite{beth31} and Hulth\'en \cite{hult38} with the help of the so-called Bethe-ansatz method. This method has been subsequently used also for a calculation of the magnetization \cite{grif64}, the elementary excitation spectrum \cite{cloi66}, the susceptibility and other thermodynamic quantities \cite{grif64}. It is noteworthy that the Bethe-ansatz method has been afterwards adapted to find similar exact solutions for the anisotropic 
spin-1/2 XXZ Heisenberg model on a linear chain as well \cite{orba58,walk59,yang66,baxt71,baxt72}. Another interesting examples of the exactly solved quantum Heisenberg models represent the antiferromagnetic spin-1/2 Heisenberg chains with the competing nearest-neighbour and the next-nearest-neighbour interactions, which are know as the so-called Majumdar-Ghosh model \cite{maju69,maju70,maju72,broe80} and the delta-chain model \cite{hama88,douc89,mont91,mont92} both having a rather peculiar dimerized 
ground state. Note furthermore that the rigorous approach, which was originally developed by 
Majumdar and Ghosh \cite{maju69} for the spin-1/2 Heisenberg chain with the competing nearest- 
and next-nearest-neighbour interactions, has been later on utilized when searching 
for an exact ground state of the antiferromagnetic spin-1/2 quantum Heisenberg model on several 
geometrically frustrated 2D and 3D lattices \cite{shas81,suth83,kant89,kuma02,kuma05,kuma08}. 
Last but not least, there also exist a rather large class of the exactly solved antiferromagnetic quantum Heisenberg models to be extended by various biquadratic and/or multispin interactions, 
which exhibit the intriguing ground state described within the valence-bond-solid picture 
\cite{aklt87,aklt88,kenn88,affl89,chay89,dmit97,soly00,bati04,cai007,tono07,asou07,grei07,rach08,kuma09}.

Finally, let us conclude our survey of the exactly solved quantum Heisenberg models 
by mentioning a few special limiting cases for which a quite general closed-form exact solution 
(not restricted only to the ground state) has been derived. Among these valuable exceptions 
one could mention the spin-1/2 quantum XY model on the linear chain \cite{lieb61,kats62,baro71} 
and the spin-1/2 Heisenberg-Ising bond alternating chain \cite{lieb61,yao02}, which have been 
exactly treated using the Jordan-Wigner fermionization approach invented by Lieb, Schultz 
and Mattis \cite{lieb61}. Apart from these two fully exactly solvable quantum spin chains, 
the quite general exact analytical solution has been also found for the spin-1/2 
quantum Heisenberg ladder with the intra-rung pair and inter-rung quartic interactions 
with the help of transfer-matrix method \cite{barr98,barr09}.

\newpage 

\section{Dual transformation}
\setcounter{equation}{0}

Some exact results for the spin-1/2 Ising model defined on 2D lattices can be obtained in a relatively
straightforward way by making use of the dual transformation \cite{kram41}. In this section, let us consider the spin-1/2 Ising model on 2D lattices in an absence of the term incorporating 
the effect of external magnetic field
\begin{eqnarray}
{\cal H} = - J \sum_{( i, j )}^{N_{\rm B}} \sigma_i \sigma_j.
\label{2d1}	
\end{eqnarray}
It is worthy to recall that $\sigma_i = \pm 1/2$ represents two-valued Ising spin variable 
located at the $i$th lattice point and the summation is restricted only to all pairs of nearest-neighbour spins. Assuming that $N$ is a total number of lattice sites and $z$ is being its coordination number 
(i.e. the number of nearest neighbours), then, there is in total $N_{\rm B}=Nz/2$ pairs 
of nearest-neighbour spins when boundary effects are neglected in the thermodynamic limit 
$N \to \infty$. Each line (bond), which connects two adjacent spins on 2D lattice, can be regarded 
as a schematic representation of the pairwise interaction $J$ between the nearest-neighbour spins. 
As usual, the central issue of our approach is to calculate the configurational partition function
\begin{eqnarray}
{\cal Z} = \sum_{\{ \sigma_i \}} \exp(- \beta {\cal H}), 
\label{2d2}	
\end{eqnarray}
where $\beta = 1/(k_{\rm B} T)$, $k_{\rm B}$ is Boltzmann's constant, $T$ is the absolute temperature and 
the suffix $\{ \sigma_i \}$ denotes a summation over all possible configurations of the Ising spins 
on a given 2D lattice. First, let us rewrite Hamiltonian (\ref{2d1}) to the form
\begin{eqnarray}
{\cal H} = - \frac{N_{\rm B} J}{4} + J \sum_{( i, j )}^{N_{\rm B}} 
\left(\frac{1}{4} - \sigma_i \sigma_j \right).
\label{2d3}	
\end{eqnarray}
Since the canonical ensemble average of Hamiltonian readily represents the internal energy, i.e. 
${\cal U} = \langle {\cal H} \rangle$, it is easy to find the following physical interpretation 
of the Hamiltonian (\ref{2d3}). Each couple of unlike oriented adjacent spins contributes to the 
sum on the right-hand-side of Eq.~(\ref{2d3}) by the energy gain $J/2$, while each couple 
of aligned adjacent spins does not contribute to this sum at all. With respect to this, 
the internal energy of a completely ordered spin system (all spins either 'up' or 'down') 
acquires its minimum value ${\cal U}= -N_{\rm B} J/4$ under the assumption $J>0$. By substituting 
the Hamiltonian (\ref{2d3}) into Eq.~(\ref{2d2}), one gets the following expression for the 
partition function 
\begin{eqnarray}
{\cal Z} = \exp \left(\frac{\beta J}{4} N_{\rm B} \right) 
           \sum_{\{ \sigma_i \}} \exp \left(- n \frac{\beta J}{2} \right), 
\label{2d4}	
\end{eqnarray}  
where $n$ stands for the total number of unaligned spin pairs within each spin configuration. 
Obviously, the sum on the right-hand-side of Eq.~(\ref{2d4}) may in principle contain just different powers of the expression $\exp(- \beta J/2)$. In the limit of zero temperature ($T \to 0$), the expression $\exp(- \beta J/2)$ tends to zero and hence, the power expansion into a series 
$\sum\limits_n \exp(- n \beta J/2)$ gives very valuable estimate of the partition function in the 
limit of sufficiently low temperatures, \textit{the low-temperature series expansion} \cite{kram41,domb49,domb74,bell99}.

\begin{figure}[t]
\begin{center}
\includegraphics[width=16cm]{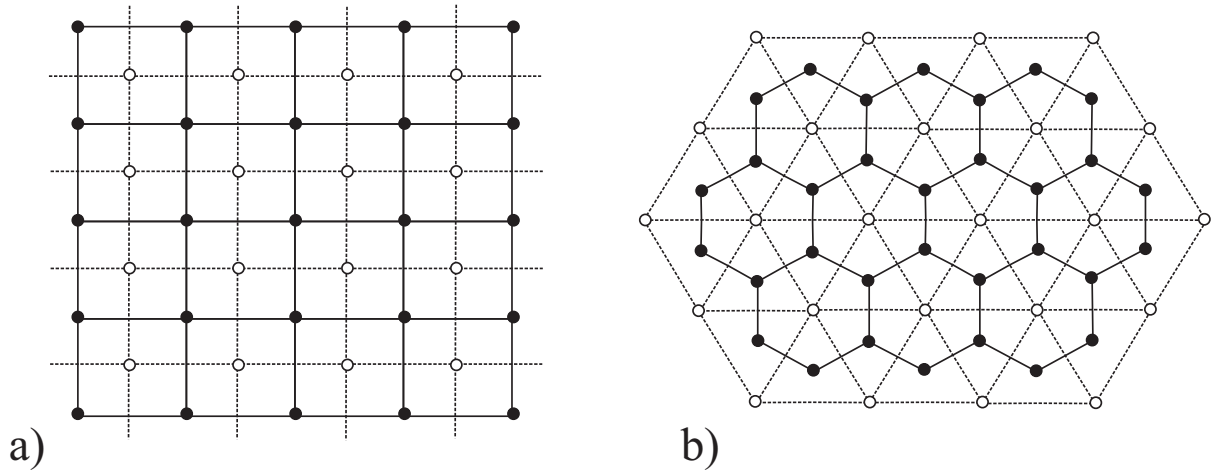}
\end{center}
\vspace{-0.8cm}
\caption{\small Two pairs of dual lattices: a) self-dual square lattices; b) honeycomb and 
triangular dual lattices. Solid lines and solid circles label edges and vertices of original 
lattices, while broken lines and empty circles stand for edges and vertices of their dual lattices, respectively.}
\label{fig:dual}
\end{figure}

It is possible to find a simple geometric interpretation of each term emerging in the 
aforementioned power series by introducing a \textit{dual lattice}. For this purpose, let us briefly
mention a basic terminology of the \textit{graph theory} \cite{essa70,temp81}, 
where each site of a lattice is called as a \textit{vertex}, while each bond (line) connecting 
the nearest-neighbour sites (vertices) is called as an \textit{edge}. Further, an interior 
of each elementary polygon delimited by edges is a \textit{face} and an ensemble of vertices 
and edges is called a \textit{full lattice graph}. Vertices of a dual lattice are simply obtained 
by situating them in the middle of each face of the original lattice. The vertices situated 
at adjacent faces, which share a common edge on the original lattice, then represent the nearest-neighbour vertices of the dual lattice. Edges of the dual lattice are obtained by connecting each couple of adjacent vertices of the dual lattice. For illustrative purposes, Fig.~\ref{fig:dual} shows two original lattices -- square and honeycomb -- and their corresponding dual lattices. 
The bonds of original 
lattices are displayed in this figure as solid lines, while the bonds of their dual lattices are depicted as broken lines. As one can see, the square lattice is a self-dual, i.e., the dual lattice 
to a square lattice is again a square lattice. Contrary to this, the triangular lattice is a dual lattice to the honeycomb lattice and vice versa. Remembering that $N$ and $N_{\rm B}$ is the total 
number of vertices and edges of the original lattice, respectively, it follows that 
$N_{\rm B}$ is at the same time the total number of edges of the dual lattice 
as well (each bond of the dual lattice intersects one and just one bond of the original lattice). 
If $N_{\rm D}$ denotes the total number of vertices of the dual lattice, then, $N_{\rm D}$ also equals to the total number of faces of the original lattice (each face of the original lattice involves 
one and just one vertex of the dual lattice). Euler's relation for planar graphs\footnote{The planar graph is a graph, which can be embedded in the plane so that no edges intersect.} consequently relates 
the total number of vertices of the original and dual lattices with the total number of edges 
\begin{eqnarray}
N + N_{\rm D} = N_{\rm B}.
\label{2d5}	
\end{eqnarray}  

\begin{figure}[t]
\begin{center}
\includegraphics[width=6cm]{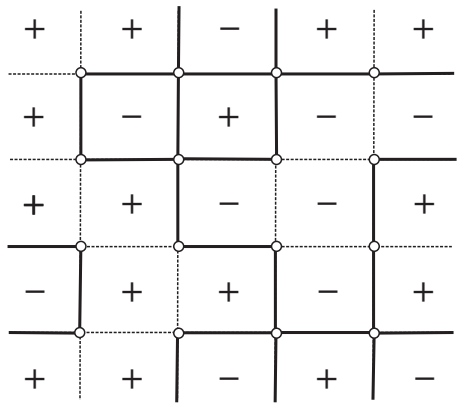}
\end{center}
\vspace{-0.8cm}
\caption{\small A particular spin configuration on a square lattice, whose vertices and edges 
are not drawn for clarity. The plus (minus) sign at $i$th lattice site corresponds to the spin 
state $\sigma_i = +1/2$ ($-1/2$). The system of solid and broken lines unambiguously determines 
the corresponding polygon line graph on the dual square lattice (for details see the text).}
\label{fig:con}
\end{figure}

Suppose now a random arrangement of 'up' and 'down' spins on the original lattice as it is displayed 
in Fig.~\ref{fig:con} for some particular example of a spin configuration on a square lattice. 
If the pairs of adjacent spins are not aligned alike, then draw solid lines between them, 
otherwise draw broken lines between them. The system of vertices, solid and broken lines 
creates a configurational graph on the dual lattice, which is a subgraph of the full 
dual lattice graph. The most fundamental property of the configurational graph is that 
the mutual interchange of each couple of adjacent spins either does not affect 
the configurational graph, or it causes an even number of changes. 
The total number of solid and broken lines incident at each site of the dual lattice 
must be therefore either even or zero. This implies that solid (broken) lines of each 
configurational graph form a system of closed polygons and hence, it follows that the 
configurational graph represents a certain kind of polygon line graph. Another important property 
of the polygon line graph is that the reversal of all spins does not change the configurational 
graph. It means that two different spin configurations (one is obtained from the other by 
reversing all spins) correspond just to one configurational polygon line graph due to 
invariance $\sigma_i \to - \sigma_i$ ($i = 1,2, \ldots, N$). The partition function of 2D 
Ising model can be therefore expressed as 
\begin{eqnarray}
{\cal Z} = 2 \exp \left(\frac{\beta J}{4} N_{\rm B} \right) 
             \sum_{{\rm p.g.}} \exp \left(- n \frac{\beta J}{2} \right), 
\label{2d6}	
\end{eqnarray}  
where the summation is now performed over all possible polygon subgraphs on the dual lattice, 
$n$ denotes the total number of solid lines within each polygon subgraph and the factor 
2 comes from the two-to-one mapping between spin and polygon configurations.

Now, let us take a closer look at another interesting property of the partition function (\ref{2d2}). 
By adopting the exact van der Waerden identity \cite{waer41}
\begin{eqnarray}
\exp(\beta J \sigma_i \sigma_j) \! \! \! &=& \! \! \!
   \cosh \left(\frac{\beta J}{4} \right) + 4 \sigma_i \sigma_j \sinh \left(\frac{\beta J}{4} \right) 
   \nonumber \\
 \! \! \! &=& \! \! \! \cosh \left(\frac{\beta J}{4} \right) 
   \left[1 + 4 \sigma_i \sigma_j \tanh \left(\frac{\beta J}{4} \right) \right] 
\label{2d7}	
\end{eqnarray}  
and substituting it into Eq.~(\ref{2d2}) one obtains
\begin{eqnarray}
{\cal Z} = \left[\cosh \left(\frac{\beta J}{4} \right) \right]^{N_{\rm B}} 
           \sum_{\{ \sigma_i \}} \prod_{(i,j)}^{N_{\rm B}} 
           \left[1 + 4 \sigma_i \sigma_j \tanh \left(\frac{\beta J}{4} \right) \right]. 
\label{2d8}	
\end{eqnarray} 
The product on the right-hand-side of Eq.~(\ref{2d8}) involves in total $N_{\rm B}$ terms, 
which give after formal multiplication a sum of in total $2^{N_{\rm B}}$ terms. However, 
many terms eventually vanish after performing a summation over all available spin configurations. 
For instance, it can be readily proved that all linear terms of the type $\sigma_i \sigma_j 
\tanh(\beta J/4)$ will disappear after summing over spin states of either the spin $\sigma_i$ 
or $\sigma_j$. The condition, which ensures that the relevant term makes a non-zero 
contribution to the partition function, can be simply guessed from the validity of the 
trivial identity $\sigma_i^2 = 1/4$. Accordingly, all non-zero terms must necessarily contain 
only spin variables, which enter into these expressions either even number of times 
or do not enter into these terms at all. The simplest non-vanishing term for the 
Ising square lattice is evidently the term
\begin{eqnarray}
4^4 (\sigma_i \sigma_j) (\sigma_j \sigma_k) (\sigma_k \sigma_l) (\sigma_l \sigma_i) 
\tanh^4 \left(\frac{\beta J}{4} \right) = 4^4 \sigma_i^2 \sigma_j^2 \sigma_k^2 \sigma_l^2 
\tanh^4 \left(\frac{\beta J}{4} \right) = \tanh^4 \left(\frac{\beta J}{4} \right), 
\nonumber
\end{eqnarray} 
which is constituted by the product of four nearest-neighbour interactions (bonds) whose corresponding
edges form an elementary square (the simplest closed polygon) on this lattice. The summation over 
spin configurations of four spins included in this term consequently gives the factor 
$2^4 \tanh^4(\beta J/4)$, while the summation over spin states of other spins yields the 
additional factor $2^{N-4}$ so that the Boltzmann factor $2^{N} \tanh^4(\beta J/4)$ is finally 
obtained as the contribution from a single square. It is therefore not difficult to construct 
the following geometric interpretation of the non-vanishing terms: edges corresponding 
to interactions to be present in these terms must create closed polygons so that either no lines 
or even number of lines meet at each vertex of 2D lattice. From this point of view, 
polygon line graphs very similar to those described by the low-temperature series expansion 
give non-zero contributions to the partition function. With regard to this, the expression 
(\ref{2d8}) for the partition function can be replaced with
\begin{eqnarray}
{\cal Z} = 2^N \left[\cosh \left(\frac{\beta J}{4} \right) \right]^{N_{\rm B}} 
           \sum_{{\rm p.g.}} \left[\tanh \left(\frac{\beta J}{4} \right) \right]^n, 
\label{2d10}	
\end{eqnarray}  
where $n$ denotes the total number of full lines constituting the particular polygon graph.
It is quite apparent that the summation on the right-hand-side of Eq.~(\ref{2d10}) contains 
just different powers of the expression $\tanh(\beta J/4)$. In the limit of high temperatures
($T \to \infty$), the expression $\tanh(\beta J/4)$ tends to zero and hence, the power expansion 
into a series $\sum\limits_n [ \tanh (\beta J/4) ]^n$ gives very valuable estimate of the partition function in the limit of high enough temperatures, \textit{the high-temperature series expansion} \cite{kram41,domb49,domb74,bell99}.

There is an interesting correspondence between summations to emerge in Eqs.~(\ref{2d6}) and (\ref{2d10}), since both of them are performed over certain sets of polygon line graphs. The most
essential difference between them lies in the fact that the summation in Eq.~(\ref{2d4}) 
is carried out over polygon graphs on the dual lattice, while the summation in Eq.~(\ref{2d10}) 
is performed over polygon graphs on the original lattice. It should be stressed, however, 
that both expressions (\ref{2d4}) and (\ref{2d10}) for the partition function are exact 
when the relevant series is performed up to an infinite order and therefore, they must 
basically give the same result for the partition function. In the thermodynamic limit, the 
summations (\ref{2d4}) and (\ref{2d10}) yield the same partition function provided that 
\begin{eqnarray}
\exp \left(- \frac{\beta_{\rm D} J}{2} \right) 
  \! \! \! &=& \! \! \! \tanh \left(\frac{\beta J}{4} \right), \label{2d11a}	\\
\frac{{\cal Z}(N_{\rm D}, \beta_{\rm D} J)}{\exp \left(\frac{\beta_{\rm D} J}{4} N_{\rm B}\right)} 
\! \! \! &=& \! \! \!  \frac{{\cal Z}(N, \beta J)}{2^{N} 
       \left[\cosh \left(\frac{\beta J}{4} \right) \right]^{N_{\rm B}}}, 
\label{2d11b}	
\end{eqnarray} 
where we have introduced the reciprocal temperature $\beta_{\rm D} = 1/(k_{\rm B} T_{\rm D})$ 
of the dual lattice and the factor $2$ was omitted from the denominator on the left-hand-side 
of Eq.~(\ref{2d11b}) as it can be neglected in the thermodynamic limit. The connection 
between two mutually dual lattices (\ref{2d11a}) and (\ref{2d11b}) can also be inverted because 
of a symmetry in the duality 
\begin{eqnarray}
\exp \left(- \frac{\beta J}{2} \right) 
  \! \! \! &=& \! \! \! \tanh \left(\frac{\beta_{\rm D} J}{4} \right), \label{2d12a}	\\
\frac{{\cal Z}(N, \beta J)}{\exp \left(\frac{\beta J}{4} N_{\rm B}\right)} 
\! \! \! &=& \! \! \!  \frac{{\cal Z}(N_{\rm D}, \beta_{\rm D} J)}{2^{N_{\rm D}} 
       \left[\cosh \left(\frac{\beta_{\rm D} J}{4} \right) \right]^{N_{\rm B}}}, 
\label{2d12b}	
\end{eqnarray} 
since each from a couple of the mutually dual lattices is dual one to another.
With the help of Eqs.~(\ref{2d11a}) and (\ref{2d11b}) [or equivalently Eqs.~(\ref{2d12a}) 
and (\ref{2d12b})], it is also possible to write this so-called \textit{dual transformation} 
even in a symmetric form as it could be expected from the symmetrical nature of the duality. 
The relation (\ref{2d11a}) gives after straightforward algebraic manipulation 
\begin{eqnarray}
\sinh \left( \frac{\beta J}{2} \right) \sinh \left( \frac{\beta_{\rm D} J}{2} \right) = 1,
\label{2d13}	
\end{eqnarray} 
while the expression (\ref{2d11b}) can be modified by regarding the equality between partition functions, Eqs.~(\ref{2d11a}), (\ref{2d13}) and Euler's relation (\ref{2d5}) implying that
\begin{eqnarray}
2^N \left[\cosh \left( \frac{\beta J}{4} \right) \right]^{N_{\rm B}} 
    \exp \left(- \frac{\beta_{\rm D} J}{4} N_{\rm B} \right) = 
     \frac{\left[2 \sinh \left( \frac{\beta J}{2} \right) \right]^{\frac{N}{2}}}
          {\left[2 \sinh \left( \frac{\beta_{\rm D} J}{2} \right) \right]^{\frac{N_{\rm D}}{2}}} = 1.
\label{2d14}	
\end{eqnarray} 
By combining Eq.~(\ref{2d14}) with the relation (\ref{2d11b}), one actually gains another symmetric relationship between the partition functions ${\cal Z}(N, \beta J)$ and 
${\cal Z}(N_{\rm D}, \beta_{\rm D} J)$, which are expressed in terms of the high-temperature series expansion on the original lattice and the low-temperature series expansion on its dual lattice
\begin{eqnarray}
\frac{{\cal Z} (N, \beta J)}{\left[2 \sinh \left( \frac{\beta J}{2} \right) \right]^{\frac{N}{2}}} 
= \frac{{\cal Z} (N_{\rm D}, \beta_{\rm D} J)}
       {\left[2 \sinh \left( \frac{\beta_{\rm D} J}{2} \right) \right]^{\frac{N_{\rm D}}{2}}}. 
\label{2d15}	
\end{eqnarray} 
An existence of the mapping equivalence between the low- and high-temperature series expansions reflects the fundamental property of the partition function, namely, its symmetry with respect 
to the low and high temperatures. This symmetry means, among other matters, that the partition function at some lower temperature can always be mapped on the equivalent partition function at some certain higher temperature. This mapping is called as the \textit{dual transformation} and the dual lattices
are actually connected one to another by means of the dual transformation. In this respect, 
the dual lattices are topological representations of the dual transformation and consequently, 
one says that the dual transformation has a character of the topological transformation. 

The mathematical formulation of the dual transformation connecting effective temperatures of 
the original and its dual lattice is represented (independently of the lattice topology) either 
by the couple of equivalent equations (\ref{2d11a}) and (\ref{2d12a}), or, respectively, by a single symmetrized relation (\ref{2d13}). The latter relationship is especially useful for a better understanding of the symmetry of the partition function with respect to the low and high temperatures. 
It is sufficient to realize that the argument of the function $\sinh(\beta J/2)$ must unavoidably decrease when the argument of the other function $\sinh(\beta_{\rm D} J/2)$ increases in order 
to preserve their constant product required by the dual transformation (\ref{2d13}). This means 
that the partition function at some lower temperature, which is obtained for instance from 
the low-temperature expansion on the dual lattice, is equivalent to the partition function 
at a certain higher temperature obtained from the high-temperature expansion on the original lattice. 
It is noteworthy that the dual transformation markedly simplifies the exact enumeration 
of thermodynamic quantities on different lattices, since 
it permits to obtain the exact solution of some quantity on arbitrary 2D lattice merely from 
the corresponding exact result for this quantity on its dual 2D lattice. For instance, the critical 
point is always accompanied with some singularity or discontinuity in the thermodynamic functions 
and this non-analyticity is always somehow reflected in the partition function as well. 
The mapping relation (\ref{2d15}) clearly shows that the partition function of the original 
lattice exhibits a non-analyticity if and only if the partition function of the dual lattice 
also has a similar non-analyticity at some corresponding temperature satisfying the duality 
relation (\ref{2d13}). Besides, the expression (\ref{2d15}) allows to calculate the partition 
function of the spin-1/2 Ising model on the one from two mutually dual lattices merely from 
the corresponding exact result for the partition function of its dual lattice model.

It is worthwhile to remark that the square lattice has an extraordinary position among 2D lattices because of its self-duality. The self-dual property together with the symmetry of the partition function with respect to the low and high temperatures is just enough for determining the critical temperature and other thermodynamic quantities precisely at a critical point. Under the assumption 
of a single critical point, the same lattice topology ensures that critical parameters must be 
equal one to another on both mutually dual square lattices. According to the dual transformation (\ref{2d13}), the critical point of the spin-1/2 Ising model on the square lattice 
must obey the condition 
\begin{eqnarray}
\sinh^2 \left( \frac{\beta_{\rm c} J}{2} \right) = 1, 
\label{2d16}	
\end{eqnarray} 
which is consistent with this value of the critical temperature $T_{\rm c}$ 
[$\beta_{\rm c} = 1/(k_{\rm B} T_{\rm c})$]
\begin{eqnarray}
\frac{k_{\rm B} T_{\rm c}}{|J|} = \frac{1}{2 \ln(1 + \sqrt{2})}.
\label{2d17}	
\end{eqnarray} 
The location of a critical point of the spin-1/2 Ising model on the square lattice, which has 
been achieved in 1941 by Kramers and Wannier \cite{kram41} with the help of the dual transformation, 
can be regarded as the first exact analytical result serving in evidence of the spontaneous 
long-range ordering. However, the main disadvantage of the dual transformation lies in the fact 
that this exact method cannot serve for calculating thermodynamic quantities out of the critical point
and moreover, the temperature symmetry of the partition function does not suffice for determining critical parameters of other 2D lattices like honeycomb and triangular lattices, which are not self-dual. 

\newpage 

\section{Algebraic transformations}
\setcounter{equation}{0}

It has been demonstrated in the preceding section that it is impossible to find from the dual transformation alone an exact critical point of the spin-1/2 Ising model on 2D lattices, which are not self-dual. This has stimulated considerable interest in the search for \textit{algebraic mapping transformations}, which would tackle this outstanding problem when combining them with the dual transformation. In what follows, our attention will be therefore focused on the most important features of algebraic transformations.  

\subsection{Star-triangle transformation}

The dual transformation (\ref{2d13}) maps the honeycomb lattice into the triangular lattice 
or vice versa and thus, it does not establish the symmetry between the low- and high-temperature 
partition function on the same lattice. It is worthy of notice that such a precise relation can be alternatively derived by making use of some algebraic mapping transformation. The star-triangle transformation, which was originally invented by Onsager in his famous work \cite{onsa44}, 
establishes this useful mapping relationship for two interesting couples of dual lattices such as honeycomb--triangular and kagom\'e--diced lattices. 

\begin{figure}[t]
\begin{center}
\includegraphics[width=15cm]{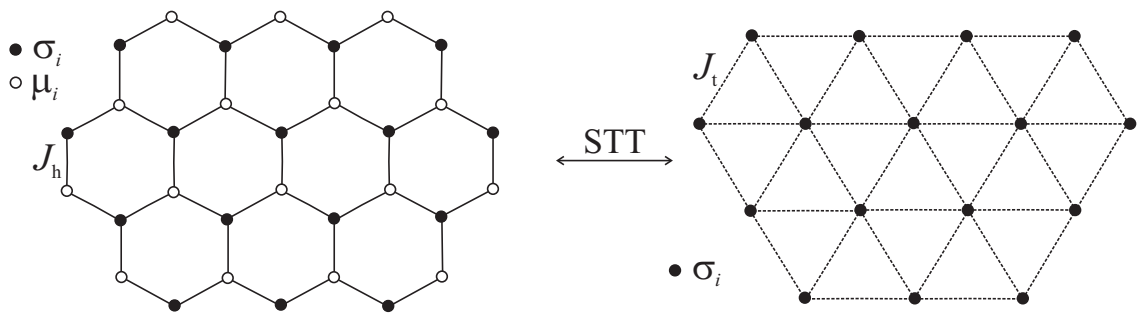}
\end{center}
\vspace{-0.8cm}
\caption{\small The spin-1/2 Ising model on the honeycomb lattice and its relation to the equivalent spin-1/2 Ising model on the triangular lattice. The equivalence between both models can be established by applying the star-triangle transformation to a half of spins of honeycomb lattice.}
\label{fig:stt}
\end{figure}

Consider the spin-1/2 Ising model on the honeycomb lattice of $2N$ sites. For further convenience, 
it is advisable to divide the honeycomb lattice into two equivalent interpenetrating triangular sublattices, whose sites are diagrammatically represented in Fig.~\ref{fig:stt} by full and empty circles, respectively. The division is made in a such way that all nearest neighbours of a site 
from the first sublattice belong to the second sublattice and vice versa. Owing to this fact, 
the summation over configurations of spins, which belong to the same sublattice, 
can be performed independently one from each other because of absence of any direct interaction 
between the spins from the same sublattice. For easy reference, let us formally denote 
the spins from the first sublattice as the \textit{vertex spins} $\sigma_i$ and the spins 
from the second sublattice as the \textit{decorating spins} $\mu_i$. The total Hamiltonian
can be for further convenience written as a sum of site Hamiltonians
\begin{eqnarray}
{\cal H} = \sum_{i = 1}^{N} {\cal H}_i, 
\label{2d18}	
\end{eqnarray} 
where each particular site Hamiltonian ${\cal H}_i$ involves all the interaction terms 
associated with the $i$th decorating spin $\mu_i$ 
\begin{eqnarray}
{\cal H}_i = -J_{\rm h} \mu_i (\sigma_{i1} + \sigma_{i2} + \sigma_{i3}). 
\label{2d19}	
\end{eqnarray} 
By the use of Eqs.~(\ref{2d18}) and (\ref{2d19}), the partition function of the spin-1/2 
Ising model on the honeycomb lattice (\ref{2d2}) can be partially factorized to the form
\begin{eqnarray}
{\cal Z}_{\rm h} = \sum_{\{ \sigma_i \}} \prod_{i = 1}^N \sum_{\mu_i = \pm 1/2} \!\!\!
\exp[\beta J_{\rm h} \mu_i (\sigma_{i1} + \sigma_{i2} + \sigma_{i3})], 
\label{2d20}	
\end{eqnarray}  
where the former summation is carried out over all available configurations of the vertex spins, the product runs over all decorating spins and the latter summation accounts for the spin states 
of one particular decorating spin. Consequently, it is adequate to consider the individual decorating spin $\mu_i$ and to sum up over degrees of freedom of this spin. It should be stressed that each decorating spin interacts merely with its three nearest-neighbour vertex spins 
and hence, this summation gives the Boltzmann's factor
\begin{eqnarray}
\sum_{\mu_i = \pm 1/2} \! \! \! \exp[\beta J_{\rm h} \mu_i (\sigma_{i1} + \sigma_{i2} + \sigma_{i3})]
\! \! \! &=& \! \! \! 2 \cosh \left[\frac{\beta J_{\rm h}}{2} (\sigma_{i1} + \sigma_{i2} + \sigma_{i3}) \right] 
\nonumber \\
\! \! \! &=& \! \! \! A \exp[\beta J_{\rm t} (\sigma_{i1} \sigma_{i2} + \sigma_{i2} \sigma_{i3} 
+ \sigma_{i3} \sigma_{i1})],
\label{2d21}	
\end{eqnarray}
which can be substituted by a simpler equivalent expression provided by the \textit{star-triangle transformation}. The physical meaning of the mapping transformation (\ref{2d21}) 
lies in removing all the interaction terms associated with the central decorating spin $\mu_{i}$  
of the \textit{star} and replacing them with some effective interactions between the three 
outer vertex spins $\sigma_{i1}$, $\sigma_{i2}$ and $\sigma_{i3}$ forming the equilateral \textit{triangle}. It is noteworthy that the star-triangle 
transformation is actually a set of eight equations, which can be obtained from the 
algebraic transformation (\ref{2d21}) by considering all possible spin configurations available 
to the three outer vertex spins. However, the consideration of eight available spin 
configurations leads just to two independent equations, which unambiguously 
determine the mapping parameters $A$ and $\beta J_{\rm t}$
\begin{eqnarray}
A \! \! \! &=& \! \! \! 2 \left[\cosh \left(\frac{3 \beta J_{\rm h}}{4} \right) \right]^{\frac{1}{4}} 
                            \left[\cosh \left(\frac{\beta J_{\rm h}}{4} \right) \right]^{\frac{3}{4}} \\ \label{2d22a}	
\beta J_{\rm t}  \! \! \! &=& \! \! \!  
\ln \left[\frac{\cosh \left(\frac{3 \beta J_{\rm h}}{4} \right)}{\cosh \left(\frac{\beta J_{\rm h}}{4} \right)}\right].
\label{2d22b}	
\end{eqnarray} 
The backward substitution of the transformation (\ref{2d21}) into the partition function (\ref{2d20}), which is equivalent with performing the star-triangle transformation for all decorating spins $\mu_i$, yields an exact mapping relationship between the partition functions of the spin-1/2 Ising model 
on the honeycomb and triangular lattices
\begin{eqnarray}
{\cal Z}_{\rm h} (2N, \beta J_{\rm h}) = A^N {\cal Z}_{\rm t} (N, \beta J_{\rm t}),
\label{2d23}	
\end{eqnarray} 
whose corresponding temperatures are coupled together through the mapping relation (\ref{2d22b}).
As a result, the mapping relation (\ref{2d22b}) resulting from the star-triangle transformation connects the partition functions of the honeycomb and triangular lattices at two different  temperatures in a very similar way as it does the relation (\ref{2d13}) provided by the dual transformation. The most crucial difference consists in a profound essence of both mapping relations; 
the dual transformation is evidently of the topological origin, whereas the star-triangle 
mapping is the algebraic transformation in its character.   

At this stage, let us combine the dual and star-triangle transformations in order to bring insight
into a criticality of the spin-1/2 Ising model on the honeycomb and triangular lattices. By 
employing a set of trivial identities for hyperbolic functions, the star-triangle transformation 
(\ref{2d22b}) can also be rewritten as follows   
\begin{eqnarray}
\exp(\beta J_{\rm t}) =  
\frac{\cosh \left(\frac{3 \beta J_{\rm h}}{4} \right)}{\cosh \left(\frac{\beta J_{\rm h}}{4} \right)}  
= 2 \cosh \left(\frac{\beta J_{\rm h}}{2} \right) - 1.
\label{2d24}	
\end{eqnarray}
Furthermore, it is appropriate to combine Eq.~(\ref{2d24}) with one of possible 
representations of the dual transformation (\ref{2d11a})
\begin{eqnarray}
\exp \left(- \frac{\beta' J_{\rm t}}{2} \right) = \tanh \left(\frac{\beta' J_{\rm h}}{4} \right)
\label{2d25}	
\end{eqnarray}
with the aim to eliminate temperature of the one from two lattices with the equivalent 
partition functions. For instance, the procedure that eliminates from Eqs.~(\ref{2d24}) and (\ref{2d25}) the effective temperature of the triangular lattice allows one to obtain the symmetrized relationship, which connects the partition function of the spin-1/2 Ising model on the honeycomb lattice at two different temperatures 
\begin{eqnarray}
\left[\cosh \left(\frac{\beta J_{\rm h}}{2} \right) - 1 \right] 
\left[\cosh \left(\frac{\beta' J_{\rm h}}{2} \right) - 1 \right] = 1.  
\label{2d26}	
\end{eqnarray}
Note that the relationship (\ref{2d26}) establishes analogous temperature symmetry for the partition function of the spin-1/2 Ising model on the honeycomb lattice as the dual transformation 
does for the spin-1/2 Ising model on the self-dual square lattice through the relation (\ref{2d13}). 
If there exists just an unique critical point, both temperatures connected via the mapping relation (\ref{2d26}) must necessarily meet at a critical point due to the same reason as it has already been explained by analysis of the square lattice. So, the critical point of the spin-1/2 Ising model on the honeycomb lattice should obey the condition
\begin{eqnarray}
\left[\cosh \left(\frac{\beta_{\rm c} J_{\rm h}}{2} \right) - 1 \right]^2 = 1,  
\label{2d27}	
\end{eqnarray}
which is consistent with this value of the critical temperature 
\begin{eqnarray}
\frac{k_{\rm B} T_{\rm c}}{|J_{\rm h}|} = \frac{1}{2 \ln(2 + \sqrt{3})}.
\label{2d28}	
\end{eqnarray}  

The same procedure can be repeated once more in order to obtain the critical parameters
of the spin-1/2 Ising model on the triangular lattice. An elimination of the effective temperature 
of the honeycomb lattice from Eqs.~(\ref{2d24}) and (\ref{2d25}) yields the following symmetric 
relationship
\begin{eqnarray}
[\exp(\beta J_{\rm t}) - 1] [\exp(\beta' J_{\rm t}) - 1] = 4,  
\label{2d29}	
\end{eqnarray}
which relates the partition function of the spin-1/2 Ising model on the triangular lattice at two different temperatures. The relation (\ref{2d29}) consecutively determines the critical condition 
for the spin-1/2 Ising model on the triangular lattice
\begin{eqnarray}
[\exp(\beta_{\rm c} J_{\rm t}) - 1]^2 = 4,  
\label{2d30}	
\end{eqnarray}
which locates its exact critical temperature
\begin{eqnarray}
\frac{k_{\rm B} T_{\rm c}}{J_{\rm t}} = \frac{1}{\ln 3}.
\label{2d31}	
\end{eqnarray}
It is noteworthy that the critical temperature of the spin-1/2 Ising model on the 
triangular lattice (\ref{2d31}) can be more easily found by substituting the critical 
condition of the spin-1/2 Ising model on the honeycomb lattice (\ref{2d27}) 
to the star-triangle transformation (\ref{2d24}). 
Of course, the same critical temperature will be recovered also in this way.

\subsection{Decoration-iteration transformation}

\begin{figure}[t]
\begin{center}
\includegraphics[width=17cm]{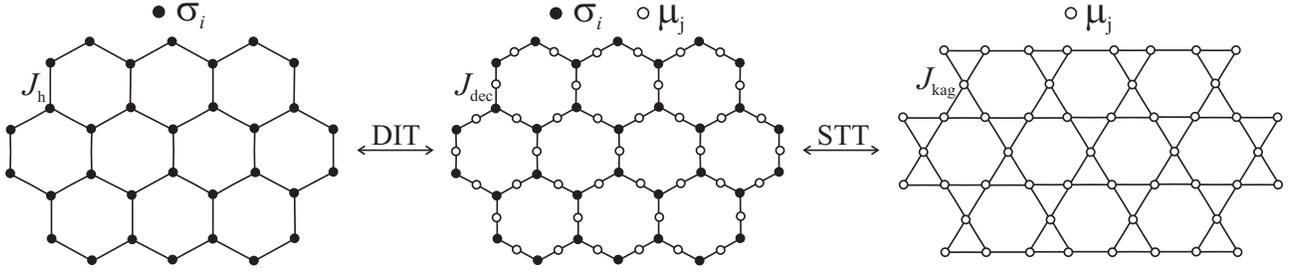}
\end{center}
\vspace{-0.8cm}
\caption{\small The spin-1/2 Ising model on the honeycomb lattice, decorated honeycomb lattice 
and kagom\'e lattice. The equivalence between all three lattice models can be established 
by employing the decoration-iteration (DIT) and star-triangle (STT) transformations, respectively.}
\label{fig:dit}
\end{figure}

Another important mapping transformation, which is purely of algebraic character, represents
the decoration-iteration transformation invented by Syozi in 1951 by solving the spin-1/2
Ising model on the kagom\'e lattice \cite{syoz51}. In principle, the approach based on 
the decoration-iteration transformation enables to obtain an exact solution of the spin-1/2 
Ising model on an arbitrary \textit{bond decorated lattice} from the corresponding exact solution 
of the simple (undecorated) lattice. The term bond decorated lattice marks such a lattice, which 
can be obtained from a simple original lattice (like square, honeycomb, triangular or any other) 
by placing an additional spin or a finite cluster of spins on the bonds of this simpler lattice. 
The spins placed at vertices of the original lattice will be referred to as \textit{vertex spins}, 
while the additional spins arising from the decoration procedure will be further called as 
\textit{decorating spins}. For illustration, Fig.~\ref{fig:dit} shows the planar topology 
of the honeycomb lattice, the simply decorated honeycomb lattice and the kagom\'e lattice 
together with the mapping relations, which can be established between them by employing 
algebraic mapping transformations. 

Let us consider initially the spin-1/2 Ising model on the decorated honeycomb lattice, 
which is shown in the middle of Fig.~\ref{fig:dit} and is given by the Hamiltonian
\begin{eqnarray}
{\cal H} = - J_{\rm d} \sum_{(i,j)}^{3N} \sigma_i \mu_j. 
\label{2d32}	
\end{eqnarray} 
Above, $\sigma_i = \pm 1/2$ and $\mu_j = \pm 1/2$ label the vertex and decorating Ising spins, respectively, the summation runs over all nearest-neighbour spin pairs on the decorated honeycomb lattice and the total number of the vertex spins is set to $N$. It is convenient to rewrite 
the total Hamiltonian (\ref{2d32}) as a sum of bond Hamiltonians
\begin{eqnarray}
{\cal H} = \sum_{j=1}^{3N/2} {\cal H}_j, 
\label{2d33}	
\end{eqnarray} 
where each bond Hamiltonian ${\cal H}_j$ involves all the interaction terms of the decorating spin $\mu_j$ from the $j$th bond of the decorated honeycomb lattice
\begin{eqnarray}
{\cal H}_j = - J_{\rm d} \mu_j (\sigma_{j1} + \sigma_{j2}). 
\label{2d34}	
\end{eqnarray} 
By the use of Eqs.~(\ref{2d33}) and (\ref{2d34}), the partition function (\ref{2d2}) of the spin-1/2 
Ising model on the decorated honeycomb lattice can be partially factorized to the form
\begin{eqnarray}
{\cal Z}_{\rm d} = \sum_{\{ \sigma_i \}} \prod_{j = 1}^{3N/2} \sum_{\mu_j = \pm 1/2} 
\! \! \! \exp[\beta J_{\rm d} \mu_j (\sigma_{j1} + \sigma_{j2})]. 
\label{2d35}	
\end{eqnarray}  
In the above expression, the former summation is carried out over all available configurations 
of the vertex spins, the product runs over all decorating spins and the latter summation accounts 
for spin states of the decorating spin $\mu_j$. According to Eq.~(\ref{2d35}), the summation over degrees of freedom of the decorating spins can be performed independently one from each other and hence, this summation gives the effective Boltzmann's factor
\begin{eqnarray}
\sum_{\mu_j = \pm 1/2} \! \! \! \! \exp[\beta J_{\rm d} \mu_j (\sigma_{j1} + \sigma_{j2})] =
2 \cosh \! \left[\frac{\beta J_{\rm d}}{2} \left(\sigma_{j1} + \sigma_{j2} \right) \right] \!
= B \exp(\beta J_{\rm h} \sigma_{j1} \sigma_{j2}),
\label{2d36}	
\end{eqnarray}
which can be successively replaced with a simpler equivalent expression provided by the  \textit{decoration-iteration transformation}. The physical meaning of the mapping transformation (\ref{2d36}) lies in removing all the interaction terms associated with the decorating spin 
$\mu_{j}$ and substituting them by the effective interaction between the two vertex spins 
$\sigma_{j1}$ and $\sigma_{j2}$, which are being its nearest neighbours. It should be emphasized 
that the decoration-iteration transformation (\ref{2d36}) has to satisfy the self-consistency condition, i.e., it must hold for any combination of the spin states of the two vertex Ising spins $\sigma_{j1}$ and $\sigma_{j2}$. In this respect, the mapping relation (\ref{2d36}) is in fact 
a set of four equations, which can be explicitly obtained by considering all possible spin configurations available to the two vertex spins $\sigma_{j1}$ and $\sigma_{j2}$. It can be readily proved that a substitution of four available spin configurations yields from the formula (\ref{2d36}) merely two independent equations, which determine so far not specified mapping parameters $B$ and $\beta J_{\rm h}$ 
\begin{eqnarray}
B \! \! \! &=& \! \! \! 2 \left[\cosh \left(\frac{\beta J_{\rm d}}{2} \right) \right]^{\frac{1}{2}}, 
\\ \label{2d37a}	
\beta J_{\rm h}  \! \! \! &=& \! \! \! 2 \ln \left[\cosh \left(\frac{\beta J_{\rm d}}{2} \right) \right].
\label{2d37b}	
\end{eqnarray} 
By applying the decoration-iteration transformation to all decorating spins, i.e.  
substituting the mapping transformation (\ref{2d36}) into the partition function (\ref{2d35}), 
one acquires an exact mapping correspondence between the partition functions 
of the spin-1/2 Ising model on the decorated honeycomb lattice and simple honeycomb lattice
\begin{eqnarray}
{\cal Z}_{\rm d} (5N/2, \beta J_{\rm dec}) = B^{3N/2} {\cal Z}_{\rm h} (N, \beta J_{\rm h}),
\label{2d38}	
\end{eqnarray} 
whose effective temperatures are connected by means of the relation (\ref{2d37b}).
It is quite obvious from Eq.~(\ref{2d38}) that the decorated honeycomb lattice becomes 
critical if and only if its corresponding honeycomb lattice becomes critical as well. 
With regard to this, it is sufficient to substitute the exact critical temperature of 
the honeycomb lattice (\ref{2d28}) into Eq.~(\ref{2d37b}) in order to locate
the critical point of the decorated honeycomb lattice
\begin{eqnarray}
\frac{k_{\rm B} T_{\rm c}}{|J_{\rm d}|} = \frac{1}{2 \ln(2 + \sqrt{3} + \sqrt{6 + 4 \sqrt{3}})}.
\label{2d39}	
\end{eqnarray} 
It is worthwhile to remark that the decoration-iteration transformation (\ref{2d36}) is not 
restricted neither by the geometry of a lattice nor by its spatial dimensionality and thus, 
it can be utilized for obtaining rigorous results for arbitrary simply decorated lattice from 
the known exact solution of the corresponding undecorated lattice.

Now, it is possible to make another useful observation. The total Hamiltonian (\ref{2d32})
of the spin-1/2 Ising model on the decorated honeycomb lattice can also be formally written
as a sum of site Hamiltonians
\begin{eqnarray}
{\cal H} = \sum_{i=1}^{N} {\cal H}_i, 
\label{2d40}	
\end{eqnarray} 
whereas each particular site Hamiltonian ${\cal H}_i$ involves all the interaction terms 
of the one individual vertex Ising spin $\sigma_i$
\begin{eqnarray}
{\cal H}_i = - J_{\rm d} \sigma_i (\mu_{i1} + \mu_{i2} + \mu_{i3}). 
\label{2d41}	
\end{eqnarray} 
Substituting Eqs.~(\ref{2d40}) and (\ref{2d41}) into a statistical definition of the partition function 
(\ref{2d2}) allows one to partially factorize the partition function of the spin-1/2 Ising model on the decorated honeycomb lattice and to rewrite it into the form
\begin{eqnarray}
{\cal Z}_{\rm d} = \sum_{\{ \mu_i \}} \prod_{i = 1}^{N} \sum_{\sigma_i = \pm 1/2} \! \! \! 
\exp[\beta J_{\rm d} \sigma_i (\mu_{i1} + \mu_{i2} + \mu_{i3})]. 
\label{2d42}	
\end{eqnarray}  
Above, the former summation runs over all available configurations of the decorating spins, 
the product runs over all vertex spins and the latter summation is carried out over the spin states 
of the $i$th vertex spin $\sigma_i$ from the decorated honeycomb lattice. The structure of the partition
function (\ref{2d42}) immediately justifies applicability of the familiar star-triangle mapping transformation
\begin{eqnarray}
\sum_{\sigma_i = \pm 1/2} \! \! \!  \exp[\beta J_{\rm d} \sigma_i (\mu_{i1} + \mu_{i2} + \mu_{i3})] \! \! \! &=& \! \! \! 2 \cosh \left[\frac{\beta J_{\rm d}}{2} (\mu_{i1} + \mu_{i2} + \mu_{i3}) \right] \nonumber \\ \! \! \! &=& \! \! \! 
C \exp[\beta J_{\rm k} (\mu_{i1} \mu_{i2} + \mu_{i2} \mu_{i3} + \mu_{i3} \mu_{i1})],
\label{2d43}	
\end{eqnarray}
which satisfies the self-consistency condition provided that
\begin{eqnarray}
C \! \! \! &=& \! \! \! 2 \left[\cosh \left(\frac{3 \beta J_{\rm d}}{4} \right) \right]^{\frac{1}{4}} 
                         \left[\cosh \left(\frac{\beta J_{\rm d}}{4} \right) \right]^{\frac{3}{4}}, 
                         \\ \label{2d44a}	
\beta J_{\rm k}  \! \! \! &=& \! \! \! 
\ln \left[\frac{\cosh \left(\frac{3 \beta J_{\rm d}}{4} \right)}{\cosh \left(\frac{\beta J_{\rm d}}{4} \right)}\right]= \ln \left[2 \cosh \left(\frac{\beta J_{\rm d}}{2} \right) - 1 \right].
\label{2d44b}	
\end{eqnarray} 
The star-triangle transformation maps the spin-1/2 Ising model on the decorated honeycomb lattice 
into the spin-1/2 Ising model on the kagom\'e lattice once it is performed for all vertex spins $\sigma_i$. 
As a matter of fact, it is easy to derive the following exact mapping equivalence between the partition functions of both these models by a direct substitution of the star-triangle transformation (\ref{2d43}) 
into the relation (\ref{2d42}) 
\begin{eqnarray}
{\cal Z}_{\rm d} (5N/2, \beta J_{\rm d}) = C^{N} {\cal Z}_{\rm k} (3N/2, \beta J_{\rm k}).
\label{2d45}	
\end{eqnarray}

At this stage, it is possible to combine the decoration-iteration transformation with the star-triangle 
transformation in order to express the partition function of the spin-1/2 Ising model on the kagom\'e lattice through the corresponding partition function of the spin-1/2 Ising model on the honeycomb lattice. The mapping relations (\ref{2d38}) and (\ref{2d45}) provide this useful connection between 
the partition functions of the spin-1/2 Ising model on the honeycomb and kagom\'e lattices
\begin{eqnarray}
{\cal Z}_{\rm k} (3N/2, \beta J_{\rm k}) = 
2^N \left\{ \frac{\exp \left(\frac{3 \beta J_{\rm h}}{2} \right)}
{\left[2 \exp \left(\frac{\beta J_{\rm h}}{2} \right) - 1 \right] 
\left[\exp \left(\frac{\beta J_{\rm h}}{2} \right) + 1 \right]^2} \right\}^{\frac{N}{4}}
        \!\!\!\!        {\cal Z}_{\rm h} (N, \beta J_{\rm h}),
\label{2d46}	
\end{eqnarray}
whereas Eqs.~(\ref{2d37b}) and (\ref{2d44b}) relate the effective temperatures of 
the honeycomb and kagom\'e lattices with the equivalent partition functions 
\begin{eqnarray}
\beta J_{\rm k} = \ln \left[ 2 \exp \left(\frac{\beta J_{\rm h}}{2} \right) - 1 \right].
\label{2d47}	
\end{eqnarray}
Substituting the exact critical temperature of the spin-1/2 Ising model on the honeycomb 
lattice (\ref{2d28}) to the mapping relation (\ref{2d47}) yields the exact critical temperature
of the spin-1/2 Ising model on the kagom\'e lattice
\begin{eqnarray}
\frac{k_{\rm B} T_{\rm c}}{J_{\rm k}} = \frac{1}{\ln (3 + 2 \sqrt{3})}.
\label{2d48}	
\end{eqnarray}

Before proceeding further, let us compare the obtained critical temperatures corresponding to the order-disorder phase transition of the spin-1/2 Ising model on several 2D lattices. For this purpose, 
the Table~\ref{table1} enumerates critical temperatures of the spin-1/2 Ising model on all three regular planar lattices -- honeycomb, square and triangular, which are the only particular plane tilings 
that entirely cover the whole plane with the same regular polygon -- hexagon, square and triangle, respectively \cite{grun87}. 
\begin{table}[t]
\begin{center}
\vspace{-0.3cm}
\vspace*{3mm}
\begin{tabular}{|c|c|c|c|c|} 
 \hline  & honeycomb & kagom\'e & square & triangular \\ \hline
 $k_{\rm B} T_{\rm c}/J$ & 0.37966 & 0.53583 & 0.56729 & 0.91024 \\ \hline 
\end{tabular} 
\end{center}
\vspace{-0.5cm}
\caption{\small Critical temperatures of the spin-1/2 Ising model on several 2D lattices.} 
\label{table1}
\end{table} 
The critical point of the semi-regular kagom\'e lattice, which consists of two kinds of regularly 
alternating polygons (hexagons and triangles) is especially interesting from the academic point 
of view, since this 2D lattice represents the only semi-regular tiling, which has all sites as well 
as all bonds equivalent quite similarly as a triad of the aforementioned regular lattices. 
It can be easily understood from the Table~\ref{table1} that the greater the coordination number 
(the number of nearest neighbours) of the planar lattice is, the higher is the critical temperature 
of its order-disorder transition. Accordingly, the cooperativity of spontaneous ordering seems
to be very closely connected with such a topological feature as the coordination number of 2D 
lattice is. On the other hand, the coordination number by itself does not entirely determine the 
critical temperature as it can be clearly seen from a comparison of the critical temperatures 
of the square and kagom\'e lattices having the same coordination number four, but slightly different 
critical temperatures. It turns out that the semi-regular Archimedean lattices \cite{grun87} 
composed of more regular polygons always have lower critical temperature than their regular counterparts with the same coordination number. 
Thus, it might be concluded that the cooperativity is strongly related also to other 
topological features of planar lattices. It should be also noted here that such an information 
cannot be elucidated from rough approximative methods, because the most of them usually predict 
the same critical temperature for 2D lattices with the same coordination number. 

\subsection{Generalized transformations I}

In two preceding parts of this section, we have shown usefulness of algebraic mapping transformations 
in providing several exact results for the spin-1/2 Ising model on 2D lattices after performing 
relatively modest calculations. From this point of view, the quite natural question arises 
whether or not algebraic transformations can be further extended and generalized. 
It is worth noticing that an early development of the concept based on \textit{generalized 
algebraic transformations} has been elaborated by Fisher in the comprehensive paper to be published more than a half century ago \cite{fish59}. In this work, Fisher has questioned a possibility 
of how the decoration-iteration transformation, the star-triangle transformation and other 
algebraic transformations can be generalized and besides, this notable paper has also furnished 
a rigorous proof on a general validity of algebraic mapping transformations. 
It should be remembered that algebraic transformations are carried out at the level 
of the partition function and their physical meaning lies in replacing a conveniently chosen 
part of the partition function with a simpler equivalent expression (Boltzmann's factor) to be obtained 
after performing a trace over degrees of freedom of a single decorating spin or a finite number 
of decorating spins, respectively. It is of principal importance that summations over spin states 
of the decorating spins (or the finite cluster of decorating spins) can be performed independently 
one from each other and before summing over spin states of the vertex Ising spins. As a result, 
the validity of generalized algebraic transformations can be verified locally by considering the Boltzmann's factor of the relevant spin cluster, which consists of the central decorating spin 
(or the finite number of decorating spins) coupled to a few outer vertex Ising spins. 

\subsubsection{Generalized decoration-iteration transformation}

\begin{figure}[t]
\begin{center}
\includegraphics[width=11cm]{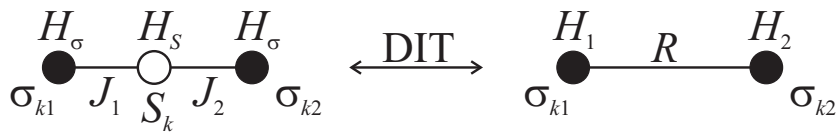}
\end{center}
\vspace{-0.7cm}
\caption{\small A diagrammatic representation of the generalized decoration-iteration transformation 
(DIT) for the decorating system composed of a single decorating spin $S_k$ of arbitrary magnitude. 
The terms $J_1$ and $J_2$ stand for the pair interactions between the decorating spin $S_k$ and its 
two nearest-neighbour vertex spins $\sigma_{k1}$ and $\sigma_{k2}$, while the terms $H_{S}$ 
and $H_{\sigma}$ represent the effect of external magnetic field acting on the decorating and vertex spins, respectively.}
\label{fig:dit1}
\end{figure}

Consider first the generalization of the decoration-iteration transformation, 
which is applicable for the models in which each decorating system interacts merely 
with the two outer vertex Ising spins. Let a single decorating Ising spin $S_{k}$ of arbitrary magnitude 
be the decorating system, which interacts with the two vertex spins $\sigma_{k1}$ 
and $\sigma_{k2}$ as it is schematically illustrated in Fig.~\ref{fig:dit1}. 
Note that the subsequent generalization to a more general decorating system composed of 
a finite number of decorating spins is quite straightforward and will be investigated hereafter. 
The generalized decoration-iteration transformation for the decorating system, which constitutes 
a single decorating Ising spin $S_{k}$ of arbitrary magnitude, can be mathematically formulated 
as follows
\begin{eqnarray}
\exp[\beta H_{\sigma} (\sigma_{k1} + \sigma_{k2})] \sum_{S_k = -S}^{S} \!\!\!\!\!\!\! && \!\!\!\!\!\!\!
\exp \left[\beta S_k \left(J_1 \sigma_{k1} + J_2 \sigma_{k2} + H_S \right) \right] \nonumber \\
\!\!\! &=& \!\!\! 
A \exp[\beta R \sigma_{k1} \sigma_{k2} + \beta H_{1} \sigma_{k1} + \beta H_{2} \sigma_{k2}].
\label{gdit1}
\end{eqnarray}
It is worthwhile to remark that the expression on the left-hand-side of the generalized decoration-iteration transformation (\ref{gdit1}) is in fact the Boltzmann's factor of the 
three-spin cluster from the left-hand-side of Fig.~\ref{fig:dit1}, which enters into the partition function of the Ising model taking into account two different pair interactions $J_1$ and $J_2$ 
between the central decorating spin and two outer vertex spins, as well as, the magnetostatic Zeeman's energy of the decorating and vertex spins in a presence of the external magnetic field of the magnitude $H_{S}$ and $H_{\sigma}$, respectively. This expression is substituted through the generalized 
mapping transformation (\ref{gdit1}) by the multiplicative factor $A$ and the Boltzmann's factor 
of the two-spin cluster from the right-hand-side of Fig.~\ref{fig:dit1}, which takes into account 
the pair interaction $R$ between the two outer vertex spins, as well as, the effect of generally non-uniform magnetic field $H_1$ and $H_2$ acting on the vertex spins $\sigma_{k1}$ and $\sigma_{k2}$, respectively. The physical meaning of the decoration-iteration transformation (\ref{gdit1}) lies in removing all the interaction terms involving the central decorating spin $S_k$ and substituting them 
through a simpler equivalent expression depending just on the two vertex Ising spins. It is noteworthy that the generalized decoration-iteration transformation holds quite generally if and only if the mapping transformation (\ref{gdit1}) remains valid regardless of four different spin configurations of the two vertex spins $\sigma_{k1}$ and $\sigma_{k2}$. This self-consistency condition determines so far not specified mapping parameters 
\begin{eqnarray}
A \! \! \! &=& \! \! \! \left(V_1 V_2 V_3 V_4 \right)^{\frac{1}{4}}, \qquad \qquad
\beta H_1 = \beta H_{\sigma} 
                    + \frac{1}{2} \ln \left( \frac{V_1 V_3}{V_2 V_4} \right), \nonumber \\
\beta R \! \! \! &=& \! \! \! \ln \left( \frac{V_1 V_2}{V_3 V_4} \right), \qquad \qquad \hspace{0.3cm}
\beta H_2 = \beta H_{\sigma} 
                    + \frac{1}{2} \ln \left( \frac{V_1 V_4}{V_2 V_3} \right), 
\label{gdit2}
\end{eqnarray}
which are expressed in terms of the functions $V_j$ $(j = 1-4)$ in order to write them 
in a more abbreviated and elegant form 
\begin{eqnarray}
V_1 \! \! \! &=& \! \! \! \! \! \sum_{n=-S}^{S} \! \! \cosh \left[ \frac{\beta n}{2} 
                          \left(J_1 + J_2 + 2 H_{S} \right) \right]\!, \quad
V_2 = \! \! \sum_{n=-S}^{S} \! \! \cosh \left[ \frac{\beta n}{2} 
                          \left(J_1 + J_2 - 2 H_{S} \right) \right]\!,  \nonumber \\ 
V_3 \! \! \! &=& \! \! \! \! \! \sum_{n=-S}^{S} \! \! \cosh \left[ \frac{\beta n}{2} 
                          \left(J_1 - J_2 + 2 H_{S} \right) \right]\!,  \quad 
V_4 = \! \! \sum_{n=-S}^{S} \! \! \cosh \left[ \frac{\beta n}{2} 
                          \left(J_1 - J_2 - 2 H_{S} \right) \right]\!.                                 \label{gdit3}
\end{eqnarray}
It should be stressed that the functions $V_j$ $(j = 1-4)$ are actually four different Boltzmann's weights, which can be obtained from the Boltzmann's factor on the left-hand-side of the 
generalized decoration-iteration transformation (\ref{gdit1}) by considering four possible 
spin configurations available to the two vertex Ising spins $\sigma_{k1}$ and $\sigma_{k2}$.

\begin{figure}[t]
\begin{center}
\includegraphics[width=11cm]{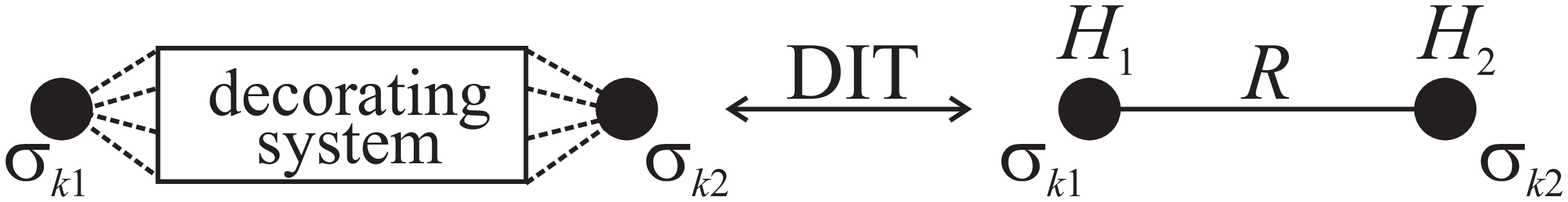}
\end{center}
\vspace{-0.6cm}
\caption{\small A diagrammatic representation of the generalized decoration-iteration transformation (DIT) for a more general decorating system composed of a finite number of decorating spins. 
Within the framework of this mapping method, the decorating system can be replaced with 
the effective pair interaction $R$ between the two outer vertex spins $\sigma_{k1}$ and $\sigma_{k2}$, as well as, the generally non-uniform magnetic fields $H_1$ and $H_2$ acting on those spins.}
\label{fig:dit2}
\end{figure}

At this stage, let us consider a more general decorating system composed of a finite 
number of the decorating spins instead of a single decorating spin as it is schematically depicted 
in Fig.~\ref{fig:dit2}. It has already been pointed out by Fisher \cite{fish59} that a single decorating spin can be in principle replaced with an arbitrary (even quantum) statistical-mechanical system without loss of the validity of the algebraic transformation. For this most general case, the generalized decoration-iteration transformation can be formally written as
\begin{eqnarray}
\mbox{Tr}_{\{S_{ki}\}} \exp \left[- \beta {\cal H}_{k} 
\left(\{S_{ki}\}, \sigma_{k1}, \sigma_{k2} \right) \right] =
A \exp(\beta R \sigma_{k1} \sigma_{k2} + \beta H_{1} \sigma_{k1} + \beta H_{2} \sigma_{k2}), 
\label{gdit4}
\end{eqnarray}
where the symbol $\mbox{Tr}_{\{S_{ki}\}}$ denotes a trace over degrees of freedom of all decorating spins and the Hamiltonian ${\cal H}_{k}$ involves all the interaction terms depending on the 
decorating spins $\{S_{ki}\}$ and the two vertex Ising spins $\sigma_{k1}$ and $\sigma_{k2}$. 
It should be also mentioned that the aforelisted expressions (\ref{gdit2}) 
for the mapping parameters will remain valid whenever the Boltzmann's factor from the left-hand-side 
of the generalized decoration-iteration transformation (\ref{gdit4}) is used for obtaining the functions $V_j$ $(j = 1-4)$ instead of those given by the set of Eqs.~(\ref{gdit3}).         

\subsubsection{Generalized star-triangle transformation}

\begin{figure}[t]
\begin{center}
\includegraphics[width=11cm]{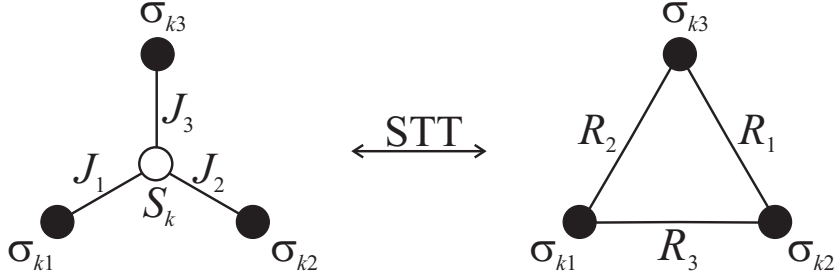}
\end{center}
\vspace{-0.6cm}
\caption{\small A diagrammatic representation of the generalized star-triangle transformation 
(STT) for the decorating system composed of a single decorating spin $S_k$ of arbitrary magnitude. 
The terms $J_1$, $J_2$ and $J_3$ stand for the pair interactions between the decorating spin $S_k$
and its three nearest-neighbour vertex spins $\sigma_{k1}$, $\sigma_{k2}$ and $\sigma_{k3}$,  respectively.}
\label{fig:stt1}
\end{figure}

Now, let us turn our attention to a generalization of the star-triangle transformation, 
which is applicable for the models in which each decorating system interacts with the three vertex Ising spins $\sigma_{k1}$, $\sigma_{k2}$ and $\sigma_{k3}$. For simplicity, we will again consider first 
the decorating system being composed of a single decorating spin $S_{k}$ of the arbitrary magnitude 
and only then proceed to the subsequent generalization dealing with a more general decorating 
system being composed of a finite number of the decorating spins. The schematic representation 
of the generalized star-triangle transformation for the particular case of the decorating system 
consisting of a single decorating spin is displayed in Fig.~\ref{fig:stt1} and this algebraic mapping 
transformation can be mathematically formulated as follows 
\begin{eqnarray}
\sum_{S_k = -S}^{S} \!\!\!\!\!\!\!\! && \!\!\!\!\!\!\!\! 
\exp \left[\beta S_k \left(J_1 \sigma_{k1}  + J_2 \sigma_{k2} + J_3 \sigma_{k2} \right) \right] 
\nonumber \\ \!\!\! &=& \!\!\! 
A \exp \left(\beta R_1 \sigma_{k2} \sigma_{k3} + \beta R_2 \sigma_{k3} \sigma_{k1} 
           + \beta R_3 \sigma_{k1} \sigma_{k2} \right).
\label{gstt1}
\end{eqnarray}
The expression on the left-hand-side of the generalized star-triangle transformation (\ref{gstt1}) 
represents the Boltzmann's factor of the four-spin cluster (star) from 
the left-hand-side of Fig.~\ref{fig:stt1}, which enters into the partition function of the Ising model taking into consideration three different pair interactions $J_1$, $J_2$ and $J_3$ between the decorating spin $S_k$ from a central site of the star and the three vertex spins $\sigma_{k1}$, $\sigma_{k2}$ and $\sigma_{k3}$ from outer sites of the star. This expression is replaced via the generalized star-triangle transformation (\ref{gstt1}) with the multiplicative factor $A$ and the Boltzmann's factor of the three-spin cluster from the right-hand-side of Fig.~\ref{fig:stt1}, which takes into account the pair interactions $R_1$, $R_2$ and $R_3$ between the three outer vertex spins $\sigma_{k1}$, $\sigma_{k2}$ and $\sigma_{k3}$ forming a triangle. The physical meaning of the star-triangle transformation (\ref{gstt1}) thus lies in removing all the interaction terms associated with the central decorating spin $S_k$  and substituting them through the effective pairwise interactions between the three outer vertex spins $\sigma_{k1}$, $\sigma_{k2}$ and $\sigma_{k3}$. 
Of course, the generalized star-triangle transformation (\ref{gstt1}) must hold independently 
of possible spin states of the three outer vertex spins and this self-consistency condition unambiguously determines the mapping parameters  
\begin{eqnarray}
A \! \! \! &=& \! \! \! \left(W_1 W_2 W_3 W_4 \right)^{\frac{1}{4}}, \qquad \qquad
\beta R_1 = \ln \left( \frac{W_1 W_4}{W_2 W_3} \right), \nonumber \\ 
\beta R_2 \! \! \! &=& \! \! \! \ln \left( \frac{W_1 W_3}{W_2 W_4} \right), \qquad \qquad \hspace{0.6cm}
\beta R_3 =  \ln \left( \frac{W_1 W_2}{W_3 W_4} \right),
\label{gstt2}
\end{eqnarray}
which are expressed for the sake of brevity using the newly defined functions 
\begin{eqnarray}
W_1 \! \! \! &=& \! \! \! \! \! \sum_{n=-S}^{S} \! \! \cosh \left[ \frac{\beta n}{2} 
                          \left(J_1 + J_2 + J_3 \right) \right]\!, \quad
W_2 = \! \! \sum_{n=-S}^{S} \! \! \cosh \left[ \frac{\beta n}{2} 
                          \left(J_1 + J_2 - J_3 \right) \right]\!,  \nonumber \\ 
W_3 \! \! \! &=& \! \! \! \! \! \sum_{n=-S}^{S} \! \! \cosh \left[ \frac{\beta n}{2} 
                          \left(J_1 - J_2 + J_3 \right) \right]\!,  \quad 
W_4 = \! \! \sum_{n=-S}^{S} \! \! \cosh \left[ \frac{\beta n}{2} 
                          \left(J_1 - J_2 - J_3 \right) \right]\!.                                 \label{gstt3}
\end{eqnarray}
Note furthermore that the functions $W_j$ $(j = 1-4)$ label four different Boltzmann's weights, 
which can be obtained from the Boltzmann's factor on the left-hand-side of the generalized star-triangle transformation (\ref{gstt1}) by considering eight available spin configurations
of the three vertex Ising spins $\sigma_{k1}$, $\sigma_{k2}$ and $\sigma_{k3}$. 

\begin{figure}[t]
\begin{center}
\includegraphics[width=11cm]{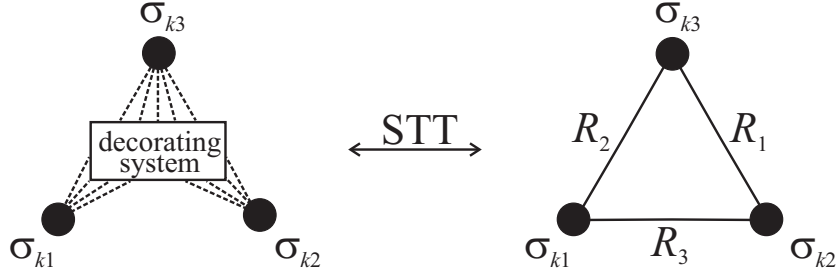}
\end{center}
\vspace{-0.6cm}
\caption{\small A diagrammatic representation of the generalized star-triangle transformation 
(STT) for arbitrary decorating system composed of a finite number of the decorating spins. 
Within the framework of this mapping method, the decorating system can be replaced with the effective pair interactions $R_1$, $R_2$ and $R_3$ between the three outer vertex Ising spins $\sigma_{k1}$, $\sigma_{k2}$ and $\sigma_{k3}$.}
\label{fig:stt2}
\end{figure}

Suppose now a more general decorating system, which comprises a finite number of the decorating spins rather than a single decorating spin, as it is schematically shown in Fig.~\ref{fig:stt2}. 
For this more general case, the generalized star-triangle transformation reads
\begin{eqnarray}
\mbox{Tr}_{\{S_{ki}\}} \!\!\!\!\!\! && \!\!\!\!\!\! \exp \left[- \beta {\cal H}_{k} 
\left(\{S_{ki}\}, \sigma_{k1}, \sigma_{k2}, \sigma_{k3} \right) \right] \nonumber \\ 
\! \! \! &=& \! \! \! 
A \exp \left(\beta R_1 \sigma_{k2} \sigma_{k3} + \beta R_2 \sigma_{k3} \sigma_{k1} 
           + \beta R_3 \sigma_{k1} \sigma_{k2} \right), 
\label{gstt4}
\end{eqnarray}
where the symbol $\mbox{Tr}_{\{S_{ki}\}}$ denotes a trace over degrees of freedom of all decorating spins and the Hamiltonian ${\cal H}_{k}$ involves all the interaction terms depending on the 
decorating spins $\{S_{ki}\}$ and the three vertex Ising spins $\sigma_{k1}$, $\sigma_{k2}$ 
and $\sigma_{k3}$. It should be mentioned that the set of expressions (\ref{gstt2}), which is 
listed above for the relevant mapping parameters, remains valid if the Boltzmann's factor from the left-hand-side of the generalized star-triangle transformation (\ref{gstt4}) is used for calculating the parameters $W_j$ $(j = 1-4)$ instead of those given by the set of Eqs.~(\ref{gstt3}).       

\subsubsection{On the validity of generalized transformations}

In this concluding part, let us briefly comment on a validity of the generalized algebraic transformations, which have been previously described in detail on two particular examples 
of the decoration-iteration and star-triangle mapping transformation. It becomes quite clear from 
a comparison of two preceding parts that the generalized decoration-iteration transformations (\ref{gdit1}) and (\ref{gdit4}) hold in general even in a presence of the non-zero external 
magnetic field, while the generalized star-triangle transformations (\ref{gstt1}) 
and (\ref{gstt4}) hold true just in an absence of the external magnetic field. 

To clarify this fundamental difference, let us at first seek a general criterion determining 
a validity of the algebraic mapping transformations in a presence of the non-zero external magnetic field. If some general decorating system interacts with the $m$ vertex Ising spins $\sigma_i$, then, the total number of mapping parameters must be either greater than or at least equal to the total number of spin configurations available to the $m$ vertex Ising spins, i.e., $2^m$. It is worth mentioning that 
one mapping parameter can always be chosen in the form of the multiplicative factor, other 
$m$ mapping parameters can be introduced by assuming the effect of inhomogeneous external 
magnetic field acting on the vertex spins and finally, $m(m-1)/2$ mapping parameters can be 
ascribed to different pair interactions between the vertex spins. The generalized algebraic transformations in a presence of the external magnetic field should therefore obey the condition
\begin{eqnarray}
1 + m + \frac{1}{2} m (m - 1) \geq 2^m,
\label{gmt1}
\end{eqnarray} 
which is satisfied if and only if the decorating system interacts with either one or two outer 
vertex Ising spins (i.e. $m \leq 2$). The aforementioned criterion thus confirms a general validity 
of the decoration-iteration mapping transformation ($m=2$) even in a presence of the non-zero external magnetic field, while the generalized star-triangle transformation ($m=3$) apparently does not satisfy 
the demanded inequality (\ref{gmt1}). 

On the other hand, the total number of independent configurations of the vertex Ising spins 
can be significantly lowered in an absence of the external magnetic field, because the relevant 
spin system has under this condition the spin inversion symmetry as it becomes invariant 
with respect to the reversal of all vertex Ising spins $\sigma_i \to - \sigma_i$ 
(for $i= 1, 2, \ldots, m$). However, it should be emphasized that the total number of mapping parameters also reduces by canceling the $m$ mapping parameters, which would represent 
the effect of inhomogeneous magnetic field acting on the vertex spins. Accordingly, the generalized algebraic transformations to be formulated in an absence of the external magnetic field should obey the inequality   
\begin{eqnarray}
1 + \frac{1}{2} m (m - 1) \geq 2^{m-1},
\label{gmt2}
\end{eqnarray}
which holds true if and only if the decorating system interacts with either one, two or three outer vertex Ising spins (i.e. $m \leq 3$). This means that the generalized decoration-iteration  
and star-triangle transformations are valid under this constraint in accordance with 
the rigorous results reported on previously. Finally, it is worthy of notice that the inequalities (\ref{gmt1}) and (\ref{gmt2}) might serve in evidence that algebraic mapping transformations for 
any decorating system coupled to four or more outer vertex Ising spins cannot hold generally 
neither in a presence nor in an absence of the external magnetic field. This assertion of course holds true provided that only the pair spin-spin and single-spin interaction terms are taken into account when constructing the generalized mapping transformations.

\subsection{Generalized transformations II}

Another important progress in the development of generalized algebraic transformations has 
recently been achieved by Rojas, Valverde and de Souza \cite{roja09} when taking into consideration 
a possibility of including higher-order multispin interactions in algebraic mapping transformations with the aim to preserve their general validity. It should be noted here that under this circumstance it is always possible to find the algebraic mapping transformation of a quite 
general validity irrespective of the number of the vertex Ising spins involved in the interactions 
with the decorating system. Before constructing the most general star-polygon mapping transformation, the rigorous approach developed by Rojas \textit{et al}. \cite{roja09} will be adopted to treat 
a quite general decorating spin system interacting with either two or three vertex Ising 
spins within the framework of the generalized decoration-iteration and star-triangle transformation, respectively.

\subsubsection{Generalized decoration-iteration transformation}

Let us consider first arbitrary decorating spin system, which may be even of a quantum nature, 
coupled to the two outer vertex Ising spins $\sigma_{k1}$ and $\sigma_{k2}$ as it is 
schematically shown in Fig.~\ref{fig:dit2}. The most general Hamiltonian of the spin 
system, which consists of all decorating spins and two vertex Ising spins, can be written as 
\begin{eqnarray}
{\cal H}_k^{(2)} \left( \{ S_{ki} \}, \sigma_{k1}, \sigma_{k2} \right) 
= \!\!\! &-& \!\!\!  J_{0,0}^{(2)} \left( \{ S_{ki} \} \right) 
  -	J_{1,0}^{(2)} \left( \{ S_{ki} \} \right) \sigma_{k1}
  -	J_{0,1}^{(2)} \left( \{ S_{ki} \} \right) \sigma_{k2} \nonumber \\
  \!\!\! &-& \!\!\!	J_{1,1}^{(2)} \left( \{ S_{ki} \} \right) \sigma_{k1} \sigma_{k2}.
\label{hamdit}
\end{eqnarray}
Here, the relevant superscript marks a total number of outer vertex spins coupled with the decorating system and the symbol $\{ S_{ki} \}$ denotes a set of all decorating spins. The physical meaning of the interaction parameters $J_{n_1, n_2}^{(2)}$ ($n_i = 0,1$ for $i = 1,2$) is quite obvious; the former 
(latter) subscript $n_1$ ($n_2$) determines whether or not the individual vertex spin $\sigma_{k1}$
($\sigma_{k2}$) is indispensable part of the relevant interaction terms. Accordingly, the parameter $J_{0,0}^{(2)}$ incorporates all the interaction terms depending only on the decorating spins 
$\{ S_{ki} \}$. Among the most common examples of this type one could mention the interaction terms representing the effect of external magnetic field and zero-field splitting parameters on the 
decorating spins, as well as, bilinear, biquadratic and other higher-order interaction terms 
between the decorating spins. Furthermore, the parameter $J_{1,0}^{(2)}$ involves all 
the interaction terms, which depend on the particular vertex spin $\sigma_{k1}$ and eventually 
on some of the decorating spins $\{S_{ki} \}$. This parameter may for instance represent the influence
of external magnetic field on the individual vertex spin $\sigma_{k1}$, as well as, the pair and higher-order interactions involving the vertex spin $\sigma_{k1}$ and some of the decorating 
spins. Similar interaction terms are also included in the parameter $J_{0,1}^{(2)}$ except 
that the vertex Ising spin $\sigma_{k2}$ now enters into the relevant interaction terms instead of $\sigma_{k1}$. Finally, the parameter $J_{1,1}^{(2)}$ involves all the interaction terms 
depending on the product $\sigma_{k1} \sigma_{k2}$ between both vertex Ising spins and eventually 
on some of the decorating spins. In this respect, the pair interaction between the two outer 
vertex spins $\sigma_{k1}$ and $\sigma_{k2}$, as well as, the higher-order multispin 
interactions incorporating both these vertex spins and some of the decorating spins, 
are possible representatives of this type.

Before presenting the most general form of the decoration-iteration transformation, it is unavoidable 
to perform a trace over degrees of freedom of all decorating spins in order to get the effective Boltzmann's factor $W_{k}^{(2)} (\sigma_{k1}, \sigma_{k2})$ depending merely on the two outer 
vertex Ising spins $\sigma_{k1}$ and $\sigma_{k2}$
\begin{eqnarray}
W_{k}^{(2)} (\sigma_{k1}, \sigma_{k2}) = 
\mbox{Tr}_{\{ S_{ki} \}} \exp \left[- \beta {\cal H}_k^{(2)} 
\left( \{ S_{ki} \}, \sigma_{k1}, \sigma_{k2} \right) \right]. 
\label{rgdit1}
\end{eqnarray}
It should be remarked that the former step, which bears a close relation to capability of finding 
the relevant trace by undertaking analytical calculation (e.g. by exact analytical diagonalization), represents perhaps the most crucial limitation to applicability of the present method to any decorating system of a quantum nature. In the latter step, the effective Boltzmann's factor (\ref{rgdit1}) can be replaced with a simpler equivalent expression depending just on the two outer vertex spins, which 
is provided by the generalized decoration-iteration mapping transformation
\begin{eqnarray}
W_{k}^{(2)} (\sigma_{k1}, \sigma_{k2})  
= \exp \left(\beta R_{0,0}^{(2)} + \beta R_{1,0}^{(2)} \sigma_{k1} + \beta R_{0,1}^{(2)} \sigma_{k2} 
           + \beta R_{1,1}^{(2)} \sigma_{k1} \sigma_{k2} \right). 
\label{rgdit2}
\end{eqnarray}
It is of great practical importance that the self-consistency condition of the generalized 
decoration-iteration transformation (\ref{rgdit2}), which ensures a general validity of this 
mapping transformation irrespective of four possible spin states of the two vertex 
Ising spins involved therein, allows one to express all four yet undetermined 
mapping parameters through the unique formula
\begin{eqnarray}
\beta R_{n_1,n_2}^{(2)} = 2^{2 (n_1 + n_2) - 2} 
\sum_{\sigma_{k1}} \sum_{\sigma_{k2}} \sigma_{k1}^{n_1} \sigma_{k2}^{n_2} 
\ln \left[ W_{k}^{(2)} (\sigma_{k1}, \sigma_{k2})\right]. 
\label{rgdit3}
\end{eqnarray}
The physical meaning of the relevant mapping parameters is as follows. The parameter $R_{1,1}^{(2)}$ 
stands for the effective pair interaction between the two outer vertex Ising spins, the parameters $R_{1,0}^{(2)}$ and $R_{0,1}^{(2)}$ represent a generally non-uniform effective magnetic field acting 
on the vertex spins $\sigma_{k1}$ and $\sigma_{k2}$, respectively, and the last parameter 
$R_{0,0}^{(2)}$ has merely a character of the multiplicative factor. With the help of the generalized decoration-iteration transformation given by Eqs. (\ref{rgdit1}) and (\ref{rgdit2}), the Hamiltonian (\ref{hamdit}) of any decorating system coupled to two outer vertex Ising spins is effectively mapped onto a very simple Hamiltonian describing two mutually interacting Ising spins 
\begin{eqnarray}
\tilde{\cal H}_k^{(2)} \left( \sigma_{k1}, \sigma_{k2} \right) 
= - R_{0,0}^{(2)} - R_{1,0}^{(2)} \sigma_{k1} - R_{0,1}^{(2)} \sigma_{k2} 
  - R_{1,1}^{(2)} \sigma_{k1} \sigma_{k2}.
\label{hamdite}
\end{eqnarray}
Of course, the temperature-dependent effective interactions $R_{n_1, n_2}^{(2)}$ of the corresponding model must obey the formula (\ref{rgdit3}) stemming from the self-consistency condition of the generalized decoration-iteration transformation in order to validate the accurate mapping equivalence 
between both Hamiltonians (\ref{hamdit}) and (\ref{hamdite}). Finally, it is also worthy to mention 
that the generalized form of the decoration-iteration transformation given by Eqs. (\ref{rgdit1}) 
and (\ref{rgdit2}) is fully consistent with the rigorous mapping transformation (\ref{gdit4}), 
which has been previously proposed following the approach due to Fisher \cite{fish59}. 

\subsubsection{Generalized star-triangle transformation}

Even more interesting situation occurs in the search for the generalized star-triangle transformation, which could be in principle applied to any decorating system coupled to 
the three outer vertex Ising spins as it is schematically illustrated in Fig.~\ref{fig:stt2}. 
It has been argued previously that the rigorous approach developed by Fisher \cite{fish59} 
does not establish the star-triangle transformation of a quite general validity due to 
a lack of the mapping parameters, which represent all possible pair spin-spin and single-spin interactions. However, the single missing mapping parameter can be chosen so as to represent 
the effective triplet interaction between the three outer vertex spins when including 
this higher-order multispin interaction into a definition of the star-triangle transformation.

The entire system, which is composed of all decorating spins and the three vertex Ising spins, 
can be described through the most general Hamiltonian of the following form
\begin{eqnarray}
{\cal H}_k^{(3)} \!\!\!\!\!\! && \!\!\!\!\!\! 
\left( \{ S_{ki} \}, \sigma_{k1}, \sigma_{k2}, \sigma_{k3} \right) =
  - J_{0,0,0}^{(3)} \left( \{ S_{ki} \} \right) 
  -	J_{1,0,0}^{(3)} \left( \{ S_{ki} \} \right) \sigma_{k1}
  -	J_{0,1,0}^{(3)} \left( \{ S_{ki} \} \right) \sigma_{k2} \nonumber \\
  \!\!\! &-& \!\!\!	J_{0,0,1}^{(3)} \left( \{ S_{ki} \} \right) \sigma_{k3}
  - J_{1,1,0}^{(3)} \left( \{ S_{ki} \} \right) \sigma_{k1} \sigma_{k2}
  - J_{0,1,1}^{(3)} \left( \{ S_{ki} \} \right) \sigma_{k2} \sigma_{k3} \nonumber \\
  \!\!\! &-& \!\!\!	J_{1,0,1}^{(3)} \left( \{ S_{ki} \} \right) \sigma_{k3} \sigma_{k1}
  - J_{1,1,1}^{(3)} \left( \{ S_{ki} \} \right) \sigma_{k1} \sigma_{k2} \sigma_{k3}. 
\label{hamstt}
\end{eqnarray}
The physical meaning of the interaction parameters $J_{n_1, n_2, n_3}^{(3)}$ ($n_i = 0,1$ for $i = 1,2,3$) is analogous as before, namely, the notation suffix $n_i$ determines whether the individual vertex spin $\sigma_{ki}$ is present ($n_i = 1$) or absent ($n_i = 0$) in the interaction terms  represented by the relevant interaction parameter. For instance, the interaction parameter $J_{1,1,1}^{(3)}$ involves all the interaction terms depending on the product 
$\sigma_{k1} \sigma_{k2} \sigma_{k3}$ of all three outer vertex Ising spins and eventually 
on some of the decorating spins $\{ S_{ki} \}$. 

After performing a trace over degrees of freedom of all decorating spins 
one obtains the effective Boltzmann's factor 
\begin{eqnarray}
W_{k}^{(3)} (\sigma_{k1}, \sigma_{k2}, \sigma_{k3}) = 
\mbox{Tr}_{\{ S_{ki} \}} \exp \left[- \beta {\cal H}_k^{(3)} 
\left( \{ S_{ki} \}, \sigma_{k1}, \sigma_{k2}, \sigma_{k3} \right) \right], 
\label{rgstt1}
\end{eqnarray}
which depends solely on the three outer vertex Ising spins $\sigma_{k1}$, $\sigma_{k2}$ 
and $\sigma_{k3}$. The Boltzmann's factor (\ref{rgstt1}) can be subsequently replaced with 
a simpler equivalent expression, which is supplied by the generalized star-triangle mapping transformation
\begin{eqnarray}
W_{k}^{(3)} (\sigma_{k1}, \sigma_{k2}, \sigma_{k3}) \!\!\! &=& \!\!\!
 \exp \Bigl(\beta R_{0,0,0}^{(3)} + \beta R_{1,0,0}^{(3)} \sigma_{k1} 
+ \beta R_{0,1,0}^{(3)} \sigma_{k2} + \beta R_{0,0,1}^{(3)} \sigma_{k3} \nonumber \\
+ \beta R_{1,1,0}^{(3)} \sigma_{k1} \sigma_{k2} 
\!\!\! &+& \!\!\! \beta R_{0,1,1}^{(3)} \sigma_{k2} \sigma_{k3} 
+ \beta R_{1,0,1}^{(3)} \sigma_{k3} \sigma_{k1}
+ \beta R_{1,1,1}^{(3)} \sigma_{k1} \sigma_{k2} \sigma_{k3}  \Bigr). 
\label{rgstt2}
\end{eqnarray}
The generalized star-triangle transformation (\ref{rgstt2}) satisfies the self-consistency condition, which demands its general validity regardless of eight available spin configurations of the three vertex Ising spins involved therein, if and only if the mapping parameters obey the unique formula
\begin{eqnarray}
\beta R_{n_1,n_2,n_3}^{(3)} = 2^{2 (n_1 + n_2 + n_3) - 3} 
\sum_{\sigma_{k1}} \sum_{\sigma_{k2}} \sum_{\sigma_{k3}} 
\sigma_{k1}^{n_1} \sigma_{k2}^{n_2} \sigma_{k3}^{n_3}
\ln \left[ W_{k}^{(3)} (\sigma_{k1}, \sigma_{k2}, \sigma_{k3}) \right]. 
\label{rgstt3}
\end{eqnarray}
The physical meaning of the relevant mapping parameters is also quite evident. The mapping parameter  $R_{0,0,0}^{(3)}$ is an appropriate multiplicative factor, while the parameters $R_{1,0,0}^{(3)}$, $R_{0,1,0}^{(3)}$ and $R_{0,0,1}^{(3)}$ represent a generally non-uniform effective magnetic field 
acting on the vertex spins $\sigma_{k1}$, $\sigma_{k2}$ and $\sigma_{k3}$, respectively. Furthermore, the mapping parameters $R_{1,1,0}^{(3)}$, $R_{0,1,1}^{(3)}$ and $R_{1,0,1}^{(3)}$ label the effective pair interactions between three exploitable couples of the vertex Ising spins and the parameter $R_{1,1,1}^{(3)}$ denotes the effective triplet interaction among them. Using the generalized star-triangle transformation 
given by Eqs. (\ref{rgstt1}) and (\ref{rgstt2}), the Hamiltonian (\ref{hamstt}) of any decorating system coupled to the three vertex Ising spins is effectively mapped onto the most general Hamiltonian of three mutually interacting Ising spins 
\begin{eqnarray}
\tilde{\cal H}_k^{(3)} \left( \sigma_{k1}, \sigma_{k2}, \sigma_{k3} \right) = 
\!\!\!&-&\!\!\! R_{0,0,0}^{(3)} - R_{1,0,0}^{(3)} \sigma_{k1} - R_{0,1,0}^{(3)} \sigma_{k2} 
  - R_{0,0,1}^{(3)} \sigma_{k3} - R_{1,1,0}^{(3)} \sigma_{k1} \sigma_{k2} \nonumber \\
\!\!\!&-&\!\!\! R_{0,1,1}^{(3)} \sigma_{k2} \sigma_{k3} - R_{1,0,1}^{(3)} \sigma_{k3} \sigma_{k1} 
  - R_{1,1,1}^{(3)} \sigma_{k1} \sigma_{k2} \sigma_{k3}.
\label{hamstte}
\end{eqnarray}
The rigorous mapping equivalence between both Hamiltonians (\ref{hamstt}) and (\ref{hamstte}) 
naturally demands that the temperature-dependent effective interactions $R_{n_1, n_2, n_3}^{(3)}$ 
of the corresponding model are in concordance with the formula (\ref{rgstt3}) derived from a self-consistency condition of the generalized star-triangle transformation. 

Finally, it should be nevertheless noted that an application of the generalized star-triangle transformation is of particular research interest just if it establishes a precise mapping 
equivalence with some simpler exactly solvable model. It is therefore valuable to mention few rigorously solved Ising models of this type. The spin-1/2 Ising model with only the triplet 
interaction is exactly tractable on the triangular lattice (the so-called Baxter-Wu model) \cite{wood72,merl72,baxtwu,fywu74,baxt74}, the union jack lattice \cite{hint72,fywu75,urumvi} 
and the diced lattice \cite{wood73,lliu74}. Among the more general exactly solved Ising models 
extended by the triplet interaction one could also mention the spin-1/2 Ising model on the 
diced \cite{hori85} and union jack \cite{urum88} lattices with the pair and triplet interactions, 
the spin-1/2 Ising model on the union jack lattice with the triplet interaction 
and the external magnetic field \cite{jung75,gitt80,urum86}, as well as, the spin-1/2 Ising 
model on the kagom\'e lattice \cite{wuwu89} and Cayley tree \cite{gani02} including 
the external magnetic field, the pair and triplet interactions.

\subsubsection{Generalized star-polygon transformation}

\begin{figure}[t]
\begin{center}
\includegraphics[width=11cm]{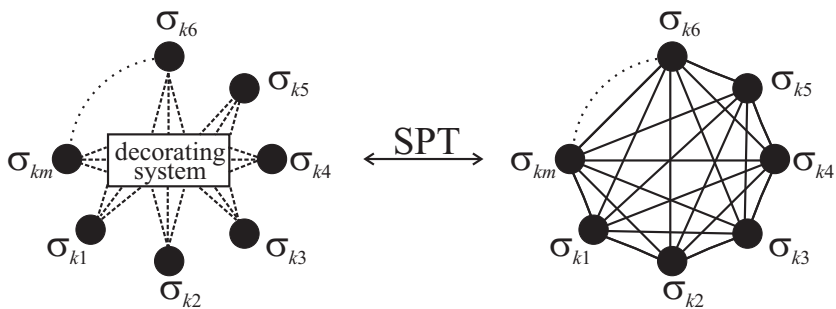}
\end{center}
\vspace{-0.7cm}
\caption{\small A diagrammatic representation of the generalized star-polygon transformation 
(SPT) for arbitrary decorating system composed of a finite number of the decorating spins. 
Within the framework of this mapping transformation, the decorating system coupled to arbitrary
number of the outer vertex Ising spins can be replaced with the effective interactions, which 
are represented through edges connecting each couple of the vertex spins, as well as, 
any polygon whose vertices all coincide with lattice points of the vertex Ising spins.}
\label{fig:spt}
\end{figure}

Last but not least, let us examine the most general form of the star-polygon transformation, 
which would enable to treat in principle any decorating system coupled to arbitrary number 
of the vertex Ising spins as it is schematically shown in Fig.~\ref{fig:spt}. It should 
be mentioned that the generalized star-polygon transformation involves the generalized decoration-iteration and star-triangle transformations as the special but surely 
the most notable cases. 
The most general Hamiltonian for the entire spin system, which consists of all decorating 
spins and the $m$ vertex Ising spins, can be written in the form
\begin{eqnarray}
{\cal H}_k^{(m)} \left( \{ S_{ki} \}, \{ \sigma_{kj} \} \right) 
= - \sum_{\{ n_j \}} J_{ \{ n_j \}}^{(m)} \left( \{ S_{ki} \} \right) 
    \prod_{j=1}^{m} \sigma_{kj}^{n_j}. 
\label{hamspt}
\end{eqnarray}
Here, the symbol $\{ n_{j} \} = \{ n_1, n_{2}, \ldots, n_{m} \}$ denotes a set of all two-valued variables, each of them determining whether the individual vertex spin $\sigma_{kj}$ is present 
($n_j = 1$) or absent ($n_j = 0$) in the interaction terms represented by the interaction parameter $J_{ \{ n_j \}}^{(m)}$. Next, the sets $\{ S_{ki} \} = \{ S_{k1}, S_{k2}, \ldots, S_{kn} \}$ and $\{ \sigma_{kj} \} = \{ \sigma_{k1}, \sigma_{k2}, \ldots, \sigma_{km} \}$ span all the decorating spins 
and vertex Ising spins, respectively, and the relevant summation $\displaystyle \sum_{\{ n_j \}} = \displaystyle \sum_{n_1 = 0,1} \displaystyle \sum_{n_2 = 0,1} \cdots \displaystyle \sum_{n_m =0,1}$ 
is carried out over the full set of two-valued variables $\{ n_{j} \}$.

The generalized star-polygon transformation can be introduced in three crucial steps. In the first step, it is necessary to perform a trace over degrees of freedom of all decorating spins in order 
to get the effective Boltzmann's factor  
\begin{eqnarray}
W_{k}^{(m)} ( \{ \sigma_{kj} \}) = \mbox{Tr}_{\{ S_{ki} \}} \exp \left[- \beta {\cal H}_k^{(m)} 
\left( \{ S_{ki} \}, \{ \sigma_{kj} \} \right) \right], 
\label{rgspt1}
\end{eqnarray}
which is solely dependent just on the $m$ vertex Ising spins from the set $\{ \sigma_{kj} \}$. 
In the second step, the Boltzmann's factor (\ref{rgspt1}) can be substituted by a simpler 
equivalent expression provided by the generalized star-polygon mapping transformation
\begin{eqnarray}
W_{k}^{(m)} ( \{ \sigma_{kj} \} ) = \exp \left[ \sum_{\{ n_j \}} \left(\beta R_{\{n_j \}}^{(m)} \prod_{j=1}^{m} \sigma_{kj}^{n_j} \right) \right]. 
\label{rgspt2}
\end{eqnarray}
In the third step, it is easy to prove that the generalized star-polygon transformation (\ref{rgspt2}) is valid independently of all spin configurations available to the $m$ vertex Ising spins if and only 
if the relevant mapping parameters fulfill the unique formula
\begin{eqnarray}
\beta R_{ \{n_j \}}^{(m)} = \frac{1}{2^m} 
\sum_{ \{\sigma_{kj} \}} \Bigl[ \prod_{j=1}^m \left(4 \sigma_{kj} \right)^{n_j} \Bigr] 
\ln \left[W_{k}^{(m)} ( \{ \sigma_{kj} \} ) \right]. 
\label{rgspt3}
\end{eqnarray}
Bearing all this in mind, the generalized star-polygon transformation given by Eqs. (\ref{rgspt1}) 
and (\ref{rgspt2}) establishes a rigorous mapping equivalence between the Hamiltonian (\ref{hamspt}) 
of any decorating system coupled to the set $\{ \sigma_{kj} \}$ of the vertex Ising spins 
and respectively, the most general Hamiltonian describing of the $m$ mutually interacting 
vertex Ising spins including all higher-order multispin interactions
\begin{eqnarray}
\tilde{\cal H}_k^{(m)} \left( \{ \sigma_{kj} \} \right) = - \sum_{\{ n_j \}}
R_{\{n_j \}}^{(m)} \prod_{j=1}^{m} \sigma_{kj}^{n_j}. 
\label{hamspte}
\end{eqnarray}
The rigorous mapping equivalence between both Hamiltonians (\ref{hamspt}) and (\ref{hamspte}) 
of course holds just if the temperature-dependent effective interactions $R_{\{n_j \}}^{(m)}$ of the corresponding model meet the unique formula (\ref{rgspt3}), which has been obtained from 
the self-consistency condition of the generalized star-polygon transformation. 

To conclude, the problem of finding a rigorous solution for any lattice-statistical model in which 
the decorating system is coupled to arbitrary number of the vertex Ising spins turns out to be equivalent 
with the problem of solving the corresponding Ising model with temperature-dependent higher-order multispin interactions between those vertex Ising spins. Another interesting observation is that 
the highest order of effective multispin interactions is in general equal to the total number 
of the vertex Ising spins to which the relevant decorating system is coupled. In the consequence 
of that, the \textit{generalized star-square transformation} as applied to any decorating system 
interacting with four outer vertex spins will for instance afford a precise mapping relationship 
with the respective Ising model taking into consideration multispin interactions up to the fourth order. From this point of view, the lattice-statistical models in which the decorating system is coupled to four or more outer vertex spins are just barely fully exactly solvable due to 
a lack of exactly solved Ising models accounting for higher-order multispin interactions. 
There are however few notable exceptions among which one could mention the spin-1/2 Ising model 
with the pair interaction on two square lattices coupled together by means of the quartic interaction, 
which is equivalent with a general eight-vertex model as first evidenced by Wu \cite{fywu71}, 
Kadanoff and Wegner \cite{kada71}. Albeit the general eight-vertex model is not exactly tractable, 
two valuable cases of the eight-vertex model has been rigorously solved under the special constraints
to its Boltzmann's weights. The symmetric eight-vertex model satisfying the zero-field condition 
has exactly been solved by Baxter \cite{zf8v71,zf8v72} and the free-fermion eight-vertex model
obeying the free-fermion condition has rigorously been solved by Fan and Wu \cite{ff8v69,ff8v70}. 
Among other matters, the utility of the generalized star-square transformation to provide exact results from a precise mapping equivalence with either the zero-field or free-fermion eight-vertex model 
will be illustrated on two particular exactly tractable models in the following two sections. 

\newpage 

\section{Exactly solved Ising models}
\setcounter{equation}{0}

In this section, let us demonstrate the undeniable capability of generalized algebraic transformations 
to enrich the realm of exactly solved Ising models. For this purpose, four scientific papers of 
the present author and collaborators are listed in the appendices A1--A4, which constitute the supplementary part of this book not included therein. These supplements are concerned 
with the Ising models, which have been exactly solved by taking advantage of the generalized decoration-iteration, star-triangle and star-square transformation, respectively. It is worth mentioning that the exactly solved Ising models, which have been earlier treated with the help 
of algebraic mapping transformations, are reviewed in the introductory parts of both papers from 
the appendices A1 and A2. In what follows, let us briefly summarize the most important scientific achievements of the papers from the appendices A1--A4, whereas the reader interested 
in more details is referred to the relevant articles for more details.   

\subsection{Decoration-iteration transformation}

In the first paper from the series on the exactly solved Ising models, which is presented as the Appendix A1, the generalized decoration-iteration transformation is employed to obtain the exact solution for the mixed-spin Ising model on a decorated square lattice with two different kinds 
of decorating spins placed on its horizontal and vertical bonds. At first sight, this exactly 
solvable model might seem as a rather trivial extension of the mixed-spin Ising model on 
a decorated square lattice with the same kind of decorating spins on all horizontal as well as 
vertical bonds, which has been proposed and exactly solved by Ja\v{s}\v{c}ur \cite{jasc98} 
and Dakhama \cite{dakh98} almost a decade ago. The most essential difference between both 
models lies in the fact that the former model with two different types of decorating spins 
is effectively mapped onto the spin-1/2 Ising model on the anisotropic square (rectangular) 
lattice, while the latter model with just one kind of decorating spins is mapped onto 
the spin-1/2 Ising model on the isotropic square lattice. 

It should be nevertheless mentioned that this more or less irrelevant difference has a very 
significant impact on the diversity of available spontaneous long-range orderings. In addition 
to the classical ferromagnetic or ferrimagnetic spin arrangement, the mixed-spin Ising model 
with the integer and half-odd-integer decorating spins placed on horizontal and 
vertical bonds of a square lattice may also exhibit a peculiar spontaneous long-range order 
of the quasi-1D character. This unusual spontaneous order appears as a result of the sufficiently strong (but not too strong) uniaxial single-ion anisotropy, which forces all integer-valued 
decorating spins towards their non-magnetic spin state. Owing to this fact, the mixed-spin 
Ising model on a decorated square lattice effectively splits into a set of quasi-independent
mixed-spin Ising chains, which comprise the alternating vertex Ising spins and half-odd-integer decorating spins. However, the most important finding to emerge from this study is that 
the investigated model surprisingly remains spontaneously long-range ordered in some 
restricted range of the uniaxial single-ion anisotropies despite the obvious effective reduction 
of its spatial dimensionality. As a matter of fact, the critical temperature tends to zero 
just at some more negative uniaxial single-ion anisotropy as one would intuitively expect 
(see Fig.~3 in the Appendix A1) and moreover, the spontaneous sublattice magnetization
of the integer-valued decorating spins is raised from zero just on behalf of selective 
thermal fluctuations (see Fig.~5(d) in the Appendix A1). These observations have obvious 
relevance to a deeper understanding of critical behaviour of quasi-1D spin systems prone 
to a spontaneous ordering below some critical temperature, which need not necessarily arise 
from much weaker interactions establishing spontaneous order 
of 3D character but may represent inherent feature of a relevant quasi-1D system instead. 
Of course, this outstanding feature cannot be observed in the mixed-spin Ising model with just 
one kind of the decorating spins, which has been explored by Ja\v{s}\v{c}ur \cite{jasc98} 
and Dakhama \cite{dakh98} in the earlier publications.  
 
\subsection{Star-triangle transformation}

The mixed spin-1/2 and spin-$S$ ($S \geq 1/2$) Ising model on a bathroom-tile lattice, which 
is the main objective of the second paper presented in the Appendix A2, has been rigorously 
treated by adopting the generalized star-triangle and triangle-star transformations. The most 
important focus of this work is to reveal the main difference between the magnetic behaviour of 
two different but relative mixed-spin Ising models on the regular honeycomb and semi-regular 
bathroom-tile lattice. It is worthwhile to remark that the honeycomb and bathroom-tile lattices 
are two planar lattices with the same coordination number three but a rather different 
lattice geometry. The regular honeycomb lattice constitute congruent polygons of only 
one type (hexagons) and accordingly, all vertices as well as all edges of this 2D lattice 
are equivalent. Contrary to this, the semi-regular bathroom-tile lattice has equivalent 
only all vertices (but not edges), because this Archimedean lattice consists of two 
different regular polygons -- squares and octagons \cite{grun87}.

It has been already mentioned before that a comparison between critical points of the spin-1/2 
Ising model on the regular square and semi-regular kagom\'e lattice (with the same coordination 
number four) has implied a slightly higher critical temperature of the former regular lattice \cite{syoz51}. 
The critical temperature apparently reflects a robustness of the spontaneous long-range order 
against thermal fluctuations and it might be therefore quite interesting to ascertain whether 
the planar lattices with the same coordination number but higher symmetry always have 
higher critical temperature than their less symmetric counterparts. Namely, it is tempting 
to conjecture that this behaviour is universal and holds true for any Ising model irrespective 
of the coordination number, spin magnitude, etc. 
The paper supplied in the Appendix A2 provides convincing evidence that the same 
conclusion is reached when comparing the respective critical points of the spin-1/2 Ising 
model on the honeycomb and bathroom-tile lattice, since the critical temperature of 
the honeycomb lattice lies just slightly above that of the bathroom-tile lattice.
Moreover, the more general mixed spin-$1/2$ and spin-$S$ Ising model on the honeycomb 
and bathroom-tile lattices also behaves completely in agreement with this assertion, 
which is evidently not affected neither by the higher spin magnitude $S \geq 1$, 
nor by the uniaxial single-ion anisotropy (see for instance Fig.~2 in the Appendix A2).

Finally, it should be stressed that another valuable by-product of calculations presented 
in the paper from the Appendix A2 is the exact solution of the spin-1/2 Ising model on the Shastry-Sutherland (orthogonal-dimer) lattice, which has been published in the subsequent separate work 
in full details \cite{stre06}. Even although the spin-1/2 Ising model on any planar lattice 
is in principle exactly tractable problem within the framework of Pfaffian technique \cite{kast61,temp61,fish61}, this rigorous method usually turns into a rather cumbersome 
and tedious calculation when applying it to some more complex planar lattice of lower symmetry. 
At the present state of knowledge, it seems quite striking that the critical temperature of the spin-1/2 Ising model has not been exactly evaluated yet for all semi-regular Archimedean lattices, 
which are the only plane tilings having all vertices equivalent \cite{grun87}. To the best of our knowledge, the Monte Carlo simulations has been implemented as the only numerical method 
in order to locate the respective critical points of the spin-1/2 Ising model on all twelve 
Archimedean lattices \cite{mala05} and thus, this analytically unresolved problem represents 
another challenging issue to deal with in the future by making use of generalized algebraic 
mapping transformations. 

\subsection{Star-square transformation}

Another two papers provided in the Appendices A3 and A4 deal with the mixed spin-1/2 
and spin-$S$ Ising model on the union jack 
(centered square) lattice, which can be rigorously mapped onto the corresponding eight-vertex model
with the help of generalized star-square transformation. It is noteworthy that the model under consideration represents an interesting example of geometrically frustrated spin system, which 
may exhibit on account of a competition between the nearest-neighbour and next-nearest-neighbour
interaction a rather rich critical behaviour including reentrant phase transitions, non-universal
critical behaviour, as well as, discontinuous first-order transitions. 

First, let us make few remarks on a critical behaviour of several particular limiting cases. 
The special case of the spin-1/2 Ising model on the union jack lattice is fully exactly solvable 
due to a precise mapping equivalence with the free-fermion eight-vertex model as convincingly 
evidenced by Wu and Lin \cite{fywu87}. In addition, this particular case simultaneously represents
the first exactly solved Ising model displaying reentrant phase transitions in a relatively 
narrow region of the parameter space \cite{vaks66}. Note furthermore that this special case 
is otherwise less interesting, since its critical behaviour belongs to the standard Ising 
universality class and besides, this model does not even exhibit first-order phase transitions. 
The much more interesting situation emerges in the particular cases with the higher spins $S \geq 1$, 
which are not completely exactly soluble on behalf of a rigorous mapping equivalence 
with the more general eight-vertex model. Exact results for those particular cases are 
therefore restricted only to a certain manifold of the parameter space, which ensures 
a precise mapping correspondence either with the zero-field \cite{zf8v71,zf8v72} or 
free-fermion \cite{ff8v69,ff8v70} eight-vertex model. In the rest of the parameter space, 
the rather meaningful and reliable results can be obtained from the free-fermion approximation 
simply ignoring a non-validity of the free-fermion condition \cite{ff8v69,ff8v70}.

It has been proved by Lipowski and Horiguchi \cite{lipo95} that the mixed spin-1/2 and spin-1 
Ising model on the union jack lattice exhibits a striking non-universal critical behaviour 
in addition to reentrant phase transitions. The non-universal criticality is closely connected 
with an appearance of special multicritical points usually referred to as \textit{bicritical points}, 
at which two second-order and one first-order phase transition lines meet together. Moreover,
Lipowski and Horiguchi \cite{lipo95} have furnished a rigorous evidence that the critical 
exponents vary continuously along the line of bicritical points in accordance 
with the weak universality hypothesis proposed by Suzuki \cite{suzu74}. It should be emphasized 
that bicritical points as well as first-order phase transitions have been calculated from a precise mapping relationship with the zero-field eight-vertex model and thus, they cannot be artefacts
of any approximation because of their exactness. The essentially same critical behaviour has 
been also detected in the mixed spin-1/2 and spin-3/2 Ising model on the union jack lattice, 
which has been examined by the present author in the publication included in the Appendix A3. 
However, the most fundamental difference between both aforementioned mixed-spin Ising models 
has been found in the allowable values of the critical exponents, which continuously vary along 
the line of bicritical points. In fact, the availability range of the critical exponents 
for the mixed-spin Ising model on the union jack lattice with $S=1$ has turned out to be 
roughly twice as large as that calculated for the analogous mixed-spin Ising model with $S=3/2$. 

To clarify this intriguing issue, another two particular cases of the mixed spin-1/2 and spin-$S$ 
Ising model on the union jack lattice with $S=2$ and $S=5/2$ have been explored in the comprehensive
work listed in the Appendix A4. This theoretical study has confirmed correctness of conclusion 
on two different ranges of availability of the critical exponents, which are governed by the 
specific condition of whether the higher spin $S$ is integer or half-odd-integer one. 

\newpage
\section{Exactly solved Ising-Heisenberg models}

In this section, let us demonstrate a versatility of the generalized algebraic 
transformations in providing exact results for the lattice-statistical models in which 
the decorating system is chosen in the form of a rather small cluster of the quantum Heisenberg spins. 
The hybrid classical-quantum spin systems composed of the semi-classical Ising spins and 
the quantum Heisenberg spins will be consequently described within the framework of \textit{Ising-Heisenberg models}, which are in principle exactly tractable by the use of 
rigorous approach developed on the grounds of generalized decoration-iteration, star-triangle 
and star-square transformations, respectively. Before proceeding to a short survey of the most 
notable findings, which are reported in detail in the papers from the Appendices B1--B4 dealing with 
three exactly solved spin-1/2 Ising-Heisenberg models, it should be pointed out that 
new exciting findings might be expected on account of a presence local 
quantum fluctuations. From this perspective, it is of particular research interest 
to explore especially 2D Ising-Heisenberg models, which might simultaneously exhibit 
a rather unusual combination of both spontaneous long-range order as well as obvious 
quantum features. 

\subsection{Decoration-iteration transformation}
\markboth{6 \hspace{0.15cm} ISING-HEISENBERG MODELS}{}

The spin-1/2 Ising-Heisenberg model on doubly decorated planar lattices, which is the main 
subject of the article from the Appendix B1, has been accurately solved by means
of the generalized decoration-iteration transformation. The most rewarding aspect of this 
rigorous study surely represents a discovery of \textit{unconventional antiferromagnetic 
spontaneous long-range order}, which occurs in spite of a pure ferromagnetic 
character of all considered interactions. It should be mentioned that this striking 
and rather unexpected quantum antiferromagnetic phase (QAP) appears as a result of the 
mutual competition between two different ferromagnetic interactions; the former is the Ising 
interaction $J_1$ of the easy-axis type and the latter one is the XXZ Heisenberg interaction 
$J (\Delta)$ of the easy-plane type. Indeed, it has been proved that the classical ferromagnetic 
order in which all Ising as well as Heisenberg spins are aligned parallel one to each other 
represents the lowest-energy state (ground state) just if the exchange anisotropy $\Delta$ 
in the XXZ Heisenberg interaction is less than the critical value $\Delta_c = \sqrt{1 + 2 J_1/J}$, 
while the unconventional QAP becomes the ground state in the reverse case $\Delta > \Delta_c \geq 1$. 

Let us take a closer look at the QAP since its mere appearance is a very surprising manifestation, 
which cannot be evidently of a classical nature but of a quantum origin. For simplicity, the unusual spin order emerging in the QAP will be described for the particular case of the spin-1/2 Ising-Heisenberg model on doubly decorated square lattice (see Fig.~1 in the Appendix B1), where it can be expressed through the eigenvector written in the following compact and suggestive form
\begin{equation}
| {\rm QAP} \rangle	=  \displaystyle \prod_{i,j} \, \Bigl| \frac{(-1)^{i+j}}{2} \Bigr \rangle \,
\prod_{{\rm {\cal D}}} \, \Bigl( \cos \phi \, \Bigl| \frac{1}{2}, - \frac{1}{2} \Bigr \rangle 
+ \, \sin \phi \, \Bigl| - \frac{1}{2}, \frac{1}{2} \Bigr \rangle \Bigr)_{{\rm {\cal D}}}. 
\label{eq:gs}
\end{equation}
Here, the mixing angle $\phi$ is given by the formula $\phi = \frac{1}{2} \arctan \left( \frac{J \Delta}{J_1} \right)$, the former product runs over all nodal sites of a square lattice occupied by the vertex Ising spins, the indices $i$ and $j$ label the relevant row and column of a square lattice, respectively, and the latter product is carried out over all bonds of a square lattice containing 
the Heisenberg spin pairs (dimers). The eigenvector (\ref{eq:gs}) indicates that the perfect 
N\'eel order is captured on the nodal sites of a square lattice by the vertex Ising spins, 
while the respective behaviour of the Heisenberg dimers is governed by the quantum entanglement 
of two antiferromagnetic microstates available to each couple of the Heisenberg spins. It should be
noticed, however, that the mutual orientation of the nearest-neighbour Ising and Heisenberg spins 
is neither perfectly parallel nor perfectly antiparallel on behalf of a zero-point motion of 
the Heisenberg spins, which is triggered by the local quantum fluctuations. According to Eq.~(\ref{eq:gs}),
the parallel (antiparallel) alignment of the nearest-neighbour Ising and Heisenberg spins 
always represents the more (less) probable microstate occurring with the probability amplitude $\cos \phi$ 
($\sin \phi$), which gradually diminishes (enhances) upon strengthening the ratio $J \Delta/J_1$.
It is noteworthy that the aforedescribed nature of the QAP is consistent with an independent analysis
of spin dynamics, which has been carried out by exploring time-dependent autocorrelation and 
correlation functions \cite{jasc04}.  

As far as the finite-temperature behaviour is concerned, the model under investigation exhibits 
quite interesting global phase diagram in which two second-order phase transition lines separating 
the classical ferromagnetic phase and the unconventional QAP from the paramagnetic region 
merge together at a quantum critical point (see Fig.~5 in the Appendix~B1). It should be also 
noted here that both critical lines are from the standard Ising universality class as convincingly 
evidenced by a detailed analysis of the relevant order parameters, which is the spontaneous magnetization 
in the classical ferromagnetic phase and the staggered magnetization in the unusual QAP. 
The same conclusion can be found from the detailed analysis of specific heat, which displays at 
a critical point of the order-disorder transition a common logarithmic singularity regardless
of whether the classical ferromagnetic or the unconventional QAP constitutes the ground state
(see Fig.~5 in the Appendix~B1). In future, it could be quite interesting to ascertain whether 
an existence of the unconventional QAP represents a generic feature of wider family of quantum 
spin models with competing easy-axis and easy-plane ferromagnetic interactions or it is just a specific
feature of the hybrid Ising-Heisenberg models. The exactly solved Ising-Heisenberg models 
could provide a useful starting basis for performing perturbative calculations, 
which could bring insight into this unresolved problem.

\subsection{Star-triangle transformation}
\markboth{6 \hspace{0.15cm} ISING-HEISENBERG MODELS}{}

The generalized star-triangle transformation has been also adapted in order to obtain the exact solution 
for the spin-1/2 Ising-Heisenberg model on the triangulated kagom\'e (triangles-in-triangles) lattice, which is supplemented in the paper listed in the Appendix~B2. It is worthy to notice that the geometric structure of the model under investigation has been motivated by the magnetic lattice of a series 
of three isostructural polymeric coordination compounds Cu$_9$X$_2$(cpa)$_6$.nH$_2$O (X = F, Cl, Br 
and cpa = carboxypentonic acid) \cite{norm87,norm90,gonz93}. The magnetic structure of this family 
of transition-metal complexes displays a rather rare and curious architecture, which is constituted 
by smaller triangular-shaped spin clusters embedded in greater triangular unit cells forming
the kagom\'e pattern \cite{norm87,norm90,gonz93}. In addition, the strong antiferromagnetic
exchange interaction between the decorating spins from smaller triangular-shaped entities 
has been evidenced by several experimental studies \cite{maru94,atec95,meka98,meka01} and thus, 
this family of magnetic compounds belongs to a prominent class of highly frustrated magnetic materials. 
In accordance with this statement, all three isostructural magnetic compounds
from this family does not order down to 1.3K despite the impressive exchange interactions of few hundreds of Kelvin \cite{maru94,atec95,meka98,meka01} and this finding might be regarded as experimental indication of a disordered spin-liquid ground state. 

With this background, the primary goal of the article presented in the Appendix~B2 has been focused 
on the role of local quantum fluctuations in determining the nature of disordered 
spin-liquid state. For this purpose, we have precisely examined both spin-1/2 Ising-Heisenberg 
model on the triangulated kagom\'e lattice as well as its analogous semi-classical Ising model 
with the aim to shed light on differences in their respective behaviour, which might be 
very pronounced due to a presence (absence) of local quantum fluctuations in the former 
(latter) model. Our exact calculations have furnished a rigorous proof that the residual entropy 
of the spin-liquid phase inherent to the Ising-Heisenberg model with any but non-zero exchange 
anisotropy amounts roughly a half of that calculated for the analogous classical Ising model. 
Owing to this fact, the local quantum fluctuations basically diminish a macroscopic degeneracy 
of the disordered spin-liquid state even though the residual entropy is not completely removed 
by the quantum 'order-from-disorder' effect unless the antisymmetric Dzyaloshinskii-Moriya 
interaction is considered \cite{cano09}.

Besides a comprehensive analysis of the disordered spin-liquid state, the critical behaviour
and basic thermodynamic quantities of the spin-1/2 Ising-Heisenberg model on the triangulated 
kagom\'e lattice have been also investigated in dependence on a strength of the geometric frustration. 
It has been evidenced that the stronger the exchange anisotropy in the XXZ Heisenberg interaction is, 
the weaker antiferromagnetic interaction is needed in order to destroy the spontaneous long-range order. In addition, the critical temperature exhibits a striking non-monotonous dependence on a strength of the geometric frustration whenever the easy-plane XXZ Heisenberg interaction competes with 
the easy-axis Ising interaction. It is worthy of notice, moreover, that our exact calculation 
has also enabled us to conjecture the notable temperature dependence of the specific heat with 
two well separated round maxima, which should be experimentally observed in the three isostructural 
polymeric coordination compounds Cu$_9$X$_2$(cpa)$_6$.nH$_2$O. 
Finally, it is worthwhile to remark that several exact results presented in the Appendix B2 
have been confirmed by a subsequent independent calculation performed by Yao \textit{et al}. \cite{dyao08}.

\subsection{Star-square transformation}
\markboth{6 \hspace{0.15cm} ISING-HEISENBERG MODELS}{}

The last two papers provided in the Appendices B3 and B4 deal with the spin-1/2 Ising-Heisenberg 
model with the pair XYZ Heisenberg interaction and the quartic Ising interaction. It should be 
pointed out that the model under investigation falls into a prominent class of fully exactly 
solved models, which are rigorously tractable due to a precise mapping equivalence with 
the zero-field eight-vertex model established through the generalized star-square transformation. 
The relevance of the exactly solved Ising-Heisenberg model with the pair Heisenberg 
interaction and the quartic Ising interaction should be viewed in more respects. The most 
attractive issue of this exactly solved model certainly represents a violation of the strong universality hypothesis, which could be intuitively expected already from a rigorous mapping 
relation with the zero-field eight-vertex model. However, the specific details of how critical exponents depend on relevant interaction parameters remains obscure until the comprehensive 
analysis is made to resolve availability range of the critical exponents. 

First, it is worthwhile to remark that the model with the antiferromagnetic pair interaction surprisingly exhibits less evident changes of both critical temperatures as well as critical exponents than the model 
with the ferromagnetic pair interaction and henceforth, our attention will be mainly concentrated on the latter model. The critical temperature of 
the spin-1/2 Ising-Heisenberg model with the ferromagnetic XXZ Heisenberg interaction displays 
a remarkable dependence on the exchange anisotropy with two marked wings of critical lines, 
which merge together at a quantum critical point between two different spontaneously long-range 
ordered phases (see Fig.~4 in the Appendix B3 and Figs.~2--4 in the Appendix B4). It should be 
emphasized that this peculiar quantum critical point is accompanied with a singular behaviour 
of the critical exponents and consequently, it bears evidence of a phase transition 
of the infinite order (see Fig.~6 in the Appendix B3 and Figs. 5--7 in the Appendix B4). 
If the XXZ ferromagnetic interaction is assumed, then, reentrant phase transitions may be also 
observed in a close vicinity of the quantum critical point due to different degeneracies 
of both spontaneously long-range ordered phases. However, the observed reentrance 
is gradually suppressed by considering the less symmetric XYZ ferromagnetic pair interaction 
as convincingly evidenced in Figs. 5--7 depicted in the Appendix B4. It has been proved 
that the disappearance of reentrant phase transitions really occurs because of the 
more general XYZ exchange anisotropy, which generally lifts a macroscopic degeneracy of the one from two spontaneously long-range ordered phases. 

To summarize, the spin-1/2 Ising-Heisenberg model with the pair XYZ Heisenberg and quartic Ising interactions represents a rare example of the exactly solved classical-quantum model with a
striking weak-universal critical behaviour. It has been demonstrated that the critical exponents 
may continuously vary over the entire range of allowable values when the relevant interaction 
parameters are appropriately chosen. Several interesting extensions and generalizations of the present 
version of the Ising-Heisenberg model come also into question. For instance, it is possible 
to extend the present model by including higher-order triplet, quintuplet and sextuplet 
interactions between the Ising and Heisenberg spins or to study thermodynamic behaviour 
of this exactly solved model in more detail. This might serve as motivation for future work. 

\newpage 

\section{Conclusions and future outlooks}
\setcounter{equation}{0}

The present book concerns with the theory of generalized algebraic transformations, which are capable of producing new interesting exactly solvable models after performing relatively modest calculations. This implies a considerable relevance of algebraic mapping transformations, since the exactly soluble models of interacting many-body systems are currently considered as an inspiring research field in its own right  \cite{domb72,thom79,baxt82,matt93,stan93,king96,lavi99,yeom02,tana02,lieb04,suth04,diep04,wu09}. 
As a matter of fact, the search for exactly solvable models is usually recognized as the highest intellectual challenge for theoretical physicists due to a rather intricate and sophisticated mathematics, which is usually involved in any attempt to treat even relatively simple lattice-statistical models rigorously \cite{lieb99,gutt05}. From this viewpoint, the rigorous method based on generalized algebraic transformations avoids those formidable mathematical difficulties by establishing a precise mapping equivalence with some simpler lattice-statistical models whose exact solutions are already known. 
Notwithstanding that this work has been aimed at promoting the implementation 
of generalized algebraic transformations, which can be adapted to discover completely novel 
and yet unexplored phenomena without a direct resemblance with any observable phenomenon of 
the corresponding exactly solved model. 

The importance of generalized algebraic transformations has been convincingly evidenced in 
two wide families of the exactly solvable models. The series of exactly solved Ising models 
brought a deeper insight into diverse aspects closely connected especially
with phase transitions and critical phenomena. More specifically, the exactly solved Ising models
from the Appendices A1--A4 shed light on a critical behaviour of spontaneously long-range ordered quasi-1D systems, reentrant phase transitions, non-universal critical behaviour, magneto-structural correlations and so on. On the other hand, the exactly solved Ising-Heisenberg 
models from the Appendices B1--B4 have served in evidence that the generalized algebraic transformations can be even adapted to treat accurately the hybrid classical-quantum models as well.
In this class of the exactly solved models, our attention has been primarily focused on 
whether spontaneous long-range order might be accompanied with obvious macroscopic 
features of a quantum origin. Among the most remarkable findings, which have been reported 
on the exactly solved Ising-Heisenberg models, one could mention the unconventional antiferromagnetic spontaneous long-range order arising from a competition between two different ferromagnetic interactions, the rigorous analysis of quantum critical points and weak-universal critical behaviour, 
as well as, the partial lifting of a macroscopic degeneracy in disordered spin liquid states 
owing to a presence of local quantum fluctuations.

Before concluding, it is worthwhile to remark that the rigorous technique based on the grounds 
of generalized algebraic transformations offers a variety of other opportunities to deal with 
in the future and the presented exactly solved models hopefully illustrate the most fundamental 
aspects of this powerful exact mapping method. It has been already argued in a breakthrough article 
published more than a half century ago by Fisher that any statistical-mechanical system 
might be selected as the decorating system without disturbing the validity of generalized algebraic transformations \cite{fish59}. Obviously, this fact implies that the generalized algebraic transformations might be employed for diverse lattice-statistical models including the ones of hybrid classical-quantum nature. Apart from the exactly solved Ising-Heisenberg models, the algebraic mapping transformations have been recently engaged also for another intriguing class of hybrid classical-quantum models describing the mutually interacting system composed of the localized Ising spins and delocalized electrons \cite{pere08,pere09,stre09,stre10}. It is the author's hope that many challenging and so far unresolved issues in the area of exactly solvable models will be tackled in the future within the framework of this rigorous mapping method.

\newpage

\markboth{Appendices}{Appendices}
\addcontentsline{toc}{section}{Appendices A1--A4}
\begin{center}
\textbf{APPENDICES A1--A4} 
\end{center}
{\small
\textbf{Appendix A1} \newline 
Exact solution of the mixed-spin Ising model on a decorated square lattice with two \\
different kinds of decorating spins on horizontal and vertical bonds \newline
Physical Review B 76 (2007) 014413 (1--9) \newline 
Jozef STRE\v{C}KA, Lucia \v{C}ANOV\'{A}, Michal JA\v{S}\v{C}UR \newline
DOI: 10.1103/PhysRevB.76.014413 \newline
\newline 
\textbf{Appendix A2} \newline 
Exact results of a mixed spin-1/2 and spin-$S$ Ising model on a bathroom-tile (4--8) lattice: \\
Effect of uniaxial single-ion anisotropy \newline
Physica A (2006) 379--390 \newline 
Jozef STRE\v{C}KA \newline
DOI: 10.1016/j.physa.2005.07.012 \newline
\newline 
\textbf{Appendix A3} \newline 
Weak universality, bicritical points and reentrant transitions in the critical behaviour of \\
a mixed spin-1/2 and spin-3/2 Ising model on the Union Jack (centered square) lattice \newline
Physica Status Solidi B 243 (2006) 708--715 \newline 
Jozef STRE\v{C}KA \newline
DOI: 10.1002/pssb.200541318 \newline
\newline 
\textbf{Appendix A4} \newline 
Weak universal critical behaviour of the mixed spin-(1/2, $S$) Ising model on the Union \\ 
Jack (centered square) lattice: integer versus half-odd-integer spin-S case \newline
Physica Status Solidi B 243 (2006) 1946--1955 \newline 
Jozef STRE\v{C}KA, Lucia \v{C}ANOV\'{A}, J\'{a}n DELY \newline
DOI: 10.1002/pssb.200642018} 

\newpage

\markboth{Appendices}{Appendices}
\addcontentsline{toc}{section}{Appendices B1--B4}
\begin{center}
\textbf{APPENDICES B1--B4} 
\end{center}
{\small
\textbf{Appendix B1} \newline 
Magnetic properties of exactly solvable doubly decorated Ising-Heisenberg planar models \newline
Physical Review B 66 (2002) 174415 (1--7) \newline 
Jozef STRE\v{C}KA, Michal JA\v{S}\v{C}UR \newline
DOI: 10.1103/PhysRevB.66.174415 \newline
\newline 
\textbf{Appendix B2} \newline 
Exact solution of the geometrically frustrated spin-1/2 Ising-Heisenberg model on the \\ triangulated kagome (triangles-in-triangles) lattice \newline
Physical Review B 78 (2008) 024427 (1--11) \newline 
Jozef STRE\v{C}KA, Lucia \v{C}ANOV\'{A}, Michal JA\v{S}\v{C}UR, Masayuki HAGIWARA \newline
DOI: 10.1103/PhysRevB.78.024427 \newline
\newline 
\textbf{Appendix B3} \newline 
Spin-1/2 Ising-Heisenberg model with the pair XYZ Heisenberg interaction and quartic \\ 
Ising interactions as the exactly soluble zero-field eight-vertex model  \newline
Physical Review E 79 (2009) 051103 (1--11) \newline
Jozef STRE\v{C}KA, Lucia \v{C}ANOV\'{A}, Kazuhiko MINAMI \newline 
DOI: 10.1103/PhysRevE.79.051103 \newline
\newline 
\textbf{Appendix B4} \newline 
Weak-universal critical behavior and quantum critical point of the exactly soluble spin-1/2 \\ Ising-Heisenberg model with the pair XYZ Heisenberg and quartic Ising interactions \newline
AIP Conference Proceedings 1198 (2009) 156--165 \newline
Jozef STRE\v{C}KA, Lucia \v{C}ANOV\'{A}, Kazuhiko MINAMI \newline
DOI: 10.1063/1.3284411}

\end{document}